\documentclass[%
preprint%
,secnumarabic%
,amssymb, amsmath,nobibnotes, aps]{revtex4}
\expandafter\ifx\csname package@font\endcsname\relax\else
 \expandafter\expandafter
 \expandafter\usepackage
 \expandafter\expandafter
 \expandafter{\csname package@font\endcsname}%
\fi

\usepackage{epsfig}%
\usepackage{graphicx}%
\usepackage{amssymb}
\usepackage{graphics}
\usepackage{amsmath}
\usepackage{amsfonts}
\usepackage{bm}
\begin{document}

\title{Neutrino coupling to cosmological background: A review on gravitational Baryo/Leptogenesis }%

\author{Gaetano Lambiase$^{a,b}$, Subhendra Mohanty$^c$, and A.R. Prasanna$^c$}%
\affiliation{$^a$ University of Salerno, Baronisi, Italy.\\
$^b$INFN , Sezione di Napoli, Italy.}
\affiliation{$^c$Physical Research Laboratory, Ahmedabad 380009, India.}
\def\be{\begin{equation}}
\def\ee{\end{equation}}
\def\al{\alpha}
\def\bea{\begin{eqnarray}}
\def\eea{\end{eqnarray}}

\begin{abstract}
In this work we review the theories of origin of matter-antimatter asymmetry in the Universe.
The general conditions for achieving baryogenesis and leptogenesis in a CPT conserving field theory have been laid down by Sakharov.
 In this review we discuss scenarios where a background scalar or gravitational field spontaneously breaks the CPT symmetry and splits the energy levels between particles and anti-particles. Baryon or Lepton number violating processes in proceeding at thermal equilibrium in such backgrounds gives rise to
  Baryon or Lepton number asymmetry.
\end{abstract}



\maketitle
\section{Introduction}

It is largely accepted that General Relativity is the best (self-consistent) theory of gravity. It dynamically describes the space-time evolution and
matter content in the Universe and is able to explain several gravitational phenomena ranging from laboratory and solar system scales (where it has been mainly tested) to astrophysical and cosmological scales. On cosmological scales, the cornerstone of General Relativity is represented by Hubble expansion, the Big Bang Nucleosynthesis, i.e. the formation of light elements in the early Universe, and cosmic microwave background (CMB) radiation. Despite these crucial predictions, Einstein's theory of gravity is in disagreement with the increasingly high number of observational data, such as those coming for example from SNIA-type, large scale structure ranging from galaxies up to galaxy super-clusters, provided by the advent of the Precision Cosmology and the achievement of high sensitivity of experiments. The experimental evidences that the observable Universe is at the moment expanding in an accelerating phase \cite{perlmutter,riess} represents without any doubts the most exciting discovery of the modern Cosmology. As a consequence of this discovery, there has been in the last years  more and more interest to understand the evolution not only of the early Universe, but also of the present Universe, and for this formidable task new ideas and theories beyond the standard Cosmology and particle physics have been proposed. Attempts to explain the recent observational data and at the same time try to preserve the conceptual structure of General Relativity, lead  cosmologists to introduce two new fundamental concepts: Dark Matter (DM) and Dark Energy (DE). Observational data indicate that a huge amounts of DM and DE are indeed needed to explain the observed cosmic acceleration of the Universe in expansion (as well as all new observational data), and at the moment there are no experimental and theoretical evidence that definitively shed some light on such mysterious components
(see \cite{copeland2006,caldwell,ellisrev03,maeda,faraonibook,defeliceLiving,faraoniRMP,nojiri,capozziello-f,capozzDE-DM,silvestri,frieman,maartens,sami}
for DE reviews and
\cite{silk,fornengo1,turner,profumo,salalti,gelmini,capozziello1,lambiase2,ahluwalia,ahluwalia1} for DM reviews, and references therein).
Moreover, recent data suggest that also in the very early epoch, the Universe was in an accelerated phase. This era is called Inflation, and is able to
to solve the problems that affect the standard cosmological model (the Cosmology based on General Relativity):
$1)$ The flatness of the Universe, that is why $\Omega=\rho/\rho_c \simeq {\cal O}(1)$. Here $\rho$ is the average cosmological energy
density, and $\rho_c=3H^2m_P^2/8\pi$ the critical density. $2)$ The problems of homogeneity, isotropy, and horizon (which created headache in the frameworks of the standard Friedman-Robertson-Walker (FRW cosmology). Inflation provides a natural mechanism of generation of
{\it small density perturbations} with almost flat spectrum\footnote{Inflation is
responsible for the inhomogeneities in the matter distribution (their evolution in fact give rise to the formation of structures, stars,
planets) and the inhomogeneities of the CMB. These perturbations are generated by {\it quantum fluctuations} of the inflaton, the scalar field that
drives the Inflation, and they can be scalar, vectorial or tensorial. The tensor perturbations allow for the primordial
gravitational waves.}. This agrees with observations.  In order to solve all problems of standard FRW cosmology, it is required that the
duration of Inflation is
 \begin{equation}\label{durationofInflation}
N\equiv H t \sim 70-100\,.
 \end{equation}
At the end of the Inflationary epoch, according to the standard cosmological model, the Universe is in a cold, low entropy
state, and appears baryon symmetric, that is the same amount of matter and antimatter.
On the contrary, the present Universe looks baryon asymmetric. The issue that one has to solve is about the physical mechanism occurred during
the Universe evolution for which it ends up being matter dominated. Equivalently, what about anti-matter?  Theories that
try to explain how the asymmetry between baryon and antibaryon was generated in the early phases of the Universe evolution are called
{\it baryogenesis}. They represent a perfect interplay between particle physics and cosmology.

The parameter characterizing the baryon asymmetry is defined as
 \[
 \eta^\prime \equiv \frac{n_B-n_{\bar B}}{n_\gamma}\,,
 \]
where $n_B$ ($n_{\bar B}$) is the number of baryons (antibaryon) per unit volume,
and $n_{\gamma}=\displaystyle{\frac{2\xi (3) T^3}{\pi^2}}$ is the photon
number density at temperature $T$. A different definition of the parameter $\eta$ that
refers to the entropy density
 \begin{equation}\label{sentropy}
 s=\frac{2\pi^2}{45}g_{eff}T^3=7.04 n_\gamma\,,
 \end{equation}
$g_{eff}=g_\gamma+\frac{T^3_\nu}{T^3_\gamma}g_\nu$,
is given by
 \[
 \eta\equiv \frac{n_B-n_{\bar B}}{s}=\frac{1}{7.04}\, \eta^\prime\,.
 \]
Finally, the baryon asymmetry can be also expressed in term of the baryonic fraction
$\Omega_B=\rho_B/\rho_c$, i.e.
 \[
\eta=2.74\times 10^{-8} \Omega_B h^2\,,
 \]
where $h=0.701\pm 0.013$ is the present Hubble parameter.

The physics of the CMB temperature  anisotropies, which are related to the acoustic oscillations of baryon-photon fluid
around the decoupling of photons, provides a strong probe of the baryon asymmetry.
In fact, the observation of the acoustic peaks in CMB measured by WMAP
satellite \cite{WMAP}, when combined with measurements of large
scale structures, leads to following estimation of the parameter $\eta$
 \begin{equation}\label{etaCMB}
 \eta^{(CMB)}\sim (6.3\pm 0.3)\times 10^{-10} \qquad \qquad 0.0215\leq \Omega_B h^2 \leq 0.0239
 \end{equation}
An independent measurement of $\eta$ can be carried out in the framework of the BBN \cite{BBN}, that gives
 \begin{equation}\label{etaBBN}
 \eta^{(BBN)}\sim (3.4-6.9)\times 10^{-10}\qquad\qquad
0.017\leq \Omega_B h^2 \leq 0.024\,.
 \end{equation}
It is remarkable that two completely different probes of the baryon content of the Universe
(the synthesis of light elements occurred during the first 3 minutes of the Universe evolution, and
the the photons decoupling occurred when the Universe was 400 thousand years old)
give compatible results. This represents one of the great success of modern Cosmology.

Although many mechanisms have been proposed, the explanation of
the asymmetry between matter and antimatter is still an open problem of the modern Cosmology and
Particle Physics.  In this review we discuss some general topics related to the baryogenesis.
The work is divided in two parts. In the first part we recall some models of baryogenesis/leptogenesis,
which are mainly based on particle physics (essentially GUT and SUSY). In the second part we discuss
different approaches to the baryon asymmetry which rely on the coupling of baryon/lepton currents with the
gravitational background. Particular attention will be devoted to the mechanism based on the spin-gravity coupling of neutrinos with the gravitational
waves of the cosmological background, which are generated by quantum fluctuations of the inflation field during the inflationary era.

\section{Some topics of Baryogenesis and Leptogenesis}
\setcounter{equation}{0}

In this Section, we recall some general topics of
Baryogenesis and Leptogenesis. More details can be found in \cite{cline,golgov1,riotto,peccei,dine,Bernreuther}.

Standard cosmological model is unable to explain the so small {\it magic} value of the baryon
asymmetry (Eqs. (\ref{etaCMB}) and (\ref{etaBBN})) and why the Universe starting from
an initial baryon symmetry ($n_B=n_{\bar B}$) evolves in  a final state such that matter dominates
over antimatter ($n_B\gg n_{\bar B}$).


\subsection{The Sakharov conditions}

As pointed out by Sakharov \cite{sakharov}, in a $CPT$ conserving theory a baryon asymmetry $B$ may be dynamically generated in the early Universe
provided that:

\begin{description}
  \item[1)] There exist interactions that violate the baryon number $B$.
Baryon number violating interactions are required because one starts from a baryon
symmetric Universe ($B=0$) to end in a baryon asymmetric Universe ($B\neq 0$).
Direct experimental proofs that baryons are not conserved
are still missing. From a theoretical point of view, both GUT and the standard electroweak theory
(via sphaleron processes) give non conservation of baryon number (notice that another possibility
to break all global charges and in particular the baryon charge is related to gravity, as discussed in
Ref. \cite{hawking}).
  \item[2)] The discrete symmetries $C$ and $CP$ must be violated. This condition is necessary in order that
matter and antimatter can be differentiated, as otherwise $B$ non-conserving interactions would produce baryons and antibaryons
at the same rate thus maintaining the net baryon number to be zero. In
contrast to non-conservation of baryons, the breaking of
$CP$-symmetry was discovered in direct experiment ($CP$ violation
has been indeed observed in the kaon system).

The $C$ and $CP$ violation imply that  $B/L$ violating reactions in
the forward and reverse channels do not cancel ($L$ stands for Lepton number).
To see this,
consider the process: $ X  \to Y +  B$, where $X$ is the initial state with $B=0$,
$Y$ the final state with $B=0$, and $B$ the excess baryon produced.
Suppose that $C$ is a symmetry. Then the $C$-conjugate process is characterized by the fact that
 \[
\Gamma( X  \to  Y  +  B)=\Gamma({\bar X}  \to { \bar Y}  + {\bar B})\,.
 \]
The net rate of baryon production evolves in time as
 \[
 \frac{dB}{dt}\propto \Gamma({\bar X}  \to { \bar Y}  +
 {\bar B})- \Gamma(X \to  Y  + B)
 =0 \quad \text{if $C$ is a symmetry}
 \]
Similar arguments hold for $CP$ symmetry. Therefore both $C$ and $CP$ discrete symmetry violation are required
to generate a net baryon asymmetry.

  \item[3)] Departure from thermal equilibrium:
This condition is required because the statistical distribution of particles and anti-particles is
the same if the Hamiltonian commutes with $CPT$, i.e.
$[H, CPT]=0$, which implies $n_B=n_{\bar B}$.
Hence only a departure from thermal equilibrium, which means that
the form of $n_{B, {\bar B}}$ has to be modified, can allow for a
finite baryon excess (so that $n_B-n_{\bar B}\neq 0$).

More specifically, consider again the process $X  \to  Y +  B $.
If the process is in {\it thermal equilibrium}, then
by definition
 \[
\Gamma( X  \to  Y  +  B )=\Gamma(Y  +  B  \to  X )\,,
 \]
 so that no net baryon asymmetry can be produced since the inverse process destroy $B$ as fast as the forward process creates it
 (see also Appendix \ref{outequ}).
\end{description}

However, as discussed in the seminal paper by Cohen and Kaplan
\cite{cohen}, it is possible to generate lepton/baryon asymmetry
at {\it thermal} equilibrium (without requiring $CP$ violation). The
reason is due to the general result that in an expanding Universe
at finite temperature, $CPT$ is not a good symmetry, i.e. $CPT$ can
be (spontaneously) violated. The third Sakharov criterion  is
therefore violated. As before pointed out, in fact, $CPT$
invariance requires that the baryon number must be generated out
of thermal equilibrium, but $CPT$ invariance requires that the
thermal distribution of baryon and anti-baryon will be identical.
This condition fails if there is a spontaneous  $CPT$ violation in the theory
which modifies the baryon-antibaryon spectrum. As a consequence one obtains
$E_{\mbox{particle}}\neq E_{\mbox{antiparticle}}$ which implies
$n_{B}-n_{\bar B} \neq 0$.

The $CPT$ violation allows for the generation of the baryon
asymmetry during an era when baryon(lepton) violating interaction are still in thermal equilibrium.
The asymmetry gets frozen at the decoupling temperature $T_d$ when the baryon(lepton) violation goes out of
equilibrium. The decoupling temperature is calculated by equating the interaction rate of processes $\Gamma$
and the expansion rate of the Universe represented by the Hubble constant $H$, $\Gamma(T_d) \simeq H(T_d)$.
The scenario underlying these processes in an expanding Universe can be schematized as follows: in the regime $\Gamma \gg H$, or $T> T_d$
B-asymmetry is generated by B-violating processes at thermal equilibrium;
at $T=T_d$, i.e. $\Gamma \simeq H$, the decoupling occurs, and finally
when $\Gamma < H$, or $T < T_d$ the B-asymmetry gets frozen.

\subsection{Leptogenesis scenario}

Leptogenesis is a mechanism, proposed by Fukugita and Yanagida, that allows to convert the lepton asymmetry to
baryon asymmetry via electroweak (EW) effects. Even if the baryon number is conserved at high scales, it is
possible to generate the baryon asymmetry in the present Universe
if lepton asymmetry is generated  at either GUT or intermediate
scales. This idea attracted much attention in view of discovery of
a possible lepton number violation in the neutrino sector
\cite{fukugita,rubakov}. For a recent review see \cite{nardi,nardi1} (the role of neutrinos in cosmology has been recently treated in \cite{miele}).

The Leptogenesis scenario is the simplest extension of the standard model able
to realize the Sakharov conditions for explaining the matter antimatter asymmetry in the Universe.
In this model, the standard model is modified by adding
right handed neutrinos which permit the implementation of the see-saw
mechanism and provide the explanation of light mass of the standard model neutrinos. At the same time,
the augmented model is able to spontaneously generate leptons from the decays of right handed neutrinos.

For later convenience, we shall discuss in a nutshell the Leptogenesis scenario.
The leptonic Lagrangian density is given by (here we follow \cite{yosefnir})
 \begin{equation}\label{LagrLeptog}
    {\cal L}= h_\beta^* ({\bar L}_\beta \, \phi^{c*}) E_\beta -\lambda_{\alpha k}^* ({\bar L}_\alpha \, \phi^{*}) N_k
    -\frac{1}{2} {\bar N}_j M_j N_j^c +h.c.
 \end{equation}
In this expression  $L$ is the Standard Model left-handed doublet, $E$ is the right-handed singlet, $N_j$ are the singlet fermions
(Majorana neutrinos), $\alpha, \beta$ are the flavor indices of the Standard Model,
i.e. $\alpha, \beta=e, \mu \tau$, $M$ is mass matrix and $\lambda$ Yakawa matrices. Equation (\ref{LagrLeptog} is written
in a basis where the coupling $h$ and the mass matrix $M$ are diagonal and real, whereas $\lambda$ is complex. This Lagrangian
leads, once the heavy fermions $N_i$ are integrated out, to the effective light neutrino masses (see-saw mechanism):
$m_{\nu\, \alpha\beta}= \lambda_{\alpha k}M^{-1}_k \lambda_{\beta k}$. The Lagrangian (\ref{LagrLeptog}) satisfies the Sakharov conditions.
In fact it violates the leptonic number $L$ due to $\lambda$- and $M$-terms; $CP$ is violated through the complex Yukawa coupling $\lambda_{\alpha k}$;
Since the interactions are only determined by Yukawa's interaction terms, the smallness of these couplings may provide the right conditions for which
the interaction rates are smaller than the expansion rate of the Universe, establishing in such a way the condition for the out of  equilibrium
(i.e. the heavy Majorana fermions can decay out of equilibrium).

For simplicity consider the lightest Majorana singlet $N_1$. It may decay in two channels
 \begin{equation}\label{Mdecay}
    N_1 \to L_{\alpha} \phi\,, \qquad N_1\to {\bar L}_\alpha \phi^\dagger\,.
 \end{equation}
As a consequence of the $N_1$ decay, the net baryon asymmetry can be generated. The parameter $\eta$ turns out to be
 \begin{equation}\label{asymmLeptog}
    \eta \simeq \frac{135 \zeta(3)}{4\pi^4 g_*} \, C_{sph} \eta_{eff} \epsilon\,.
 \end{equation}
Here $\eta_{eff}$ is the efficiency factor which assumes the value in the range $0< \eta_{eff} < 1$ (owing to
inverse decays, washout processes and inefficiency in $N_1$ production). Below to the free-out temperature $T_F$, the temperature
for which $\Gamma(\phi L\to N_1) < H$, where $\Gamma(\phi L\to N_1)\simeq \frac{1}{2}\Gamma_D e^{-M_1/T}$, with
 \begin{equation}\label{GammaD}
 \Gamma_D=\frac{(\lambda^\dagger \lambda)_{11}M_1}{8\pi}\,,
 \end{equation}
and $H$ is the expansion rate of the Universe during the radiation dominated era,
\begin{equation}\label{exprateH}
H=1.66 g_*^{1/2}\, \frac{T^2}{M_P^2}\,,
 \end{equation}
the density of the fermion $N_1$ is Boltzmann suppressed ($N_1 \sim e^{-M_1/T}$). Therefore, below $T_F$ the decay of $N_1$ contribute to the lepton asymmetry,
and the efficiency factor is
 \begin{equation}\label{effciencyfactor}
    \eta_{eff}\simeq \frac{n_{N_1}(T_F)}{n_{N_1}(T\gg M_1)}\simeq e^{-M_1/T_F}\simeq \frac{m_*}{{\tilde m}}\,, \quad {\tilde m}< m_*\,,
 \end{equation}
where
 \begin{equation}\label{m*mtilde}
    {\tilde m}\equiv \frac{8\pi v^2}{M_1^2}\Gamma_D = \frac{(\lambda^\dagger \lambda)_{11}v^2}{M_1}\,, \qquad
    m_* \equiv \frac{8\pi v^2}{M_1^2} H(T=M_1) \simeq 1.1 \times 10^{-3}\text{eV}\,.
 \end{equation}
$C_{sph}$ is a factor that takes into account  the dilution of the asymmetry due to fast processes. Finally $\epsilon$ is the $CP$ parameter related to the asymmetry in the $N_1$ decays and defined as
 \begin{equation}\label{epsilonLepto}
    \epsilon = \frac{ \Gamma(N_1\to \phi L)-  \Gamma(N_1\to \phi^\dagger {\bar L})}{ \Gamma(N_1\to \phi L) + \Gamma(N_1\to \phi^\dagger {\bar L})}\,.
 \end{equation}
Its non vanishing value arises from the interference of three level and one loop amplitudes (complex Yakawa couplings). One gets
 \begin{equation}\label{28}
    \epsilon = \frac{1}{8\pi} \frac{1}{(\lambda^\dagger \lambda)_{11}}\sum_j {\cal Im}\left\{\left[ (\lambda^\dagger \lambda)_{1j}\right]^2\right\} g(x_j)\,,
 \end{equation}
where
 \[
 g(x) = \sqrt{x}\left[\frac{2-x}{1-x} -(1+x) \ln \frac{1+x}{x}\right]\,, \qquad x_j \equiv \frac{M_j^2}{M_1^2}\,.
 \]
Taking into account Eqs. (\ref{effciencyfactor}), (\ref{28}) and (\ref{m*mtilde}), the net lepton asymmetry is
 \begin{equation}\label{asymmLeptog1}
    \eta \simeq 10^{-3} \frac{10^{-3}\text{eV}}{{\tilde m}} \epsilon\,.
 \end{equation}
Leptogenesis is then related to Baryogenesis by a phenomenon that happens in the currently accepted Standard Model.
Indeed, certain non-perturbative configurations of gauge fields, the sphalerons, can convert leptons into baryons and vice versa.
These processes that violate $B+L$ and conserve $B-L$ occur at the electroweak scale.
Under normal conditions sphalerons processes are unobservably rare due to the fact that the transition rates are extremely small
$\Gamma\sim e^{-16\pi^2/g^2}\sim {\cal O}(10^{-165})$, hence are completely negligible in the Standard Model (at $T=0$). However as
emphasized by Kuzmin, Rubakov, and Shaposhnikov \cite{rubakov}, in the thermal bah provided by the expanding Universe, thermal
fluctuations becomes important and $B+L$ violating processes can occur at a significant rate and these processes can be in
equilibrium in the expanding Universe (see \ref{SphPhysics}).

Finally from (\ref{asymmLeptog1}) it follows that by requiring $\eta\sim 10^{-10}$ the lower bound of the mass of the Majorana neutrinos $N_1$ is
 \begin{equation}\label{M1lowebouond}
    M_1 \gtrsim 10^{11} \text{GeV}\,,
 \end{equation}
where the light value of the neutrino mass has been used: ${\tilde m}\simeq (10^{-3}- 10^{-1})$eV.

\subsection{Models of Baryogenesis}

Many models aimed to explain the
generation of the baryon asymmetry have been proposed in literature. These are GUT Baryogenesis,
Affleck-Dine Baryogenesis and Affleck-Dine Leptogenesis,
Leptogenesis from heavy Majorana neutrinos, Leptogensis from
$\nu_R$ oscillation, Thermal baryogenesis, Electroweak
baryogenesis, Spontaneous baryogenesis, Baryogenesis through
evaporation of primordial black holes. Details of such mechanisms
can be found in \cite{cline,golgov1,riotto,peccei,dine,fukugitabook}
and references therein. Here we list some of them in Table \ref{ta1}.

\begin{itemize}

  \item GUT-Baryogenesis or decay of heavy particles:

Consider the $X$-boson decays in two channels $X\to 2q$ and $X\to 2{\bar q}$,
with the probabilities given by
$P_{X\to 2q}\neq P_{X\to 2{\bar q}}$ due to CP violation.
This implies the excess of baryons over anti-baryons. In the original scenario of GUT
baryogenesis one uses the heavy gauge bosons $X$ and $Y$
(leptoquarks), which decay while they decouple from equilibrium.
This is called {\it delayed decay scenario}. It was soon
realized that this boson gauge decay does not produce the
required baryon asymmetry because that the $X$ and $Y$ boson masses predicted are too low to satisfy the out-of-equilibrium
condition (in non-SUSY GUT). The alternative scenario was to use
decays of coloured Higgs particles. If more than two Higgs particles exist, sufficiently large baryon asymmetry can be
generated (provided that the Kuzmin, Rubakov, Shaposhnikov effect is switched off).

\item SUSY: Supersymmetry actually opens a number
  of options. Since Supersymmetry extends the particle content of
  the theory near the EW scale, the possibility of a strong EW
  phase transition cannot yet be completely excluded. This revives
  the hope of explaining baryon asymmetry entirely within the
  MSSM.

   {\it Affleck-Dine scenario} (1985): This scenario is based on the observation that in SUSY theories ordinary
  quarks and leptons are accompanied by supersymmetric partners -  s-quarks and s-leptons - which are scalars.
  The corresponding scalar fields carry baryon and lepton number, which can in principle be very large in the case of a
  scalar condensate (classical scalar field). An important feature of SUSY theories is the existence of flat directions
  in the superpotential, along which the relevant components of the complex scalar fields $\varphi$ can be considered as massless. The
  condensate is frozen until supersymmetry breaking takes place. Supersymmetry breaking lifts the flat directions and the scalar
  fields acquire mass. When the Hubble constant becomes of the order of this mass, the scalar fields starts to oscillate and
  decays. At this time, $B$, $L$, and $CP$ violating terms  (for example, quartic couplings
  $\lambda_1 \varphi^3\varphi^*+c.c.$ and $\lambda_2 \varphi^4+c.c.$, with complex $\lambda_{1,2}$) becomes important and a
  substantial baryon asymmetry can be produced. The scalar particles decay into ordinary quarks and leptons transferring to
  them the generated baryon asymmetry.

  The Affleck-Dine mechanism can be implemented at nearly any energy scale, even below 200 GeV. By suitable choice of the
  parameters one can explain almost any amount of baryon asymmetry and this lack of a falsifiable prediction  is an unattractive feature of the Affleck-Dine mechanism.


  \item Electroweak baryogenesis ($\sim TeV$): The asymmetry is generated by phase transitions involving $SU(2)\times U(1)$
  breaking. The EWBG is assumed to occur during the radiation dominated era of the early Universe, a period in which the $SU(2)_L\times U(1)_Y$ electroweak symmetry is manifest. As the temperature falls down the EW scale ($T_{EW}\sim 100$GeV), the Higgs field acquires an expectation value and the electroweak
symmetry is spontaneously broken to the subgroup $U(1)$. The EWBG occurs during this phase transition, and in order that it could be an available mechanism,
it is required that the transition is of the first order. Remarkably the EWBG satisfies all the three Sakharov's conditions: 1) The rapid sphaleron transitions in the symmetric phase provide the required violation of the baryon number; 2) The scattering of plasma with bubble walls generates the C and CP asymmetry of the number of particles if the underlying theory does contain terms that violate these discrete symmetries (these processes bias the sphalerons to create more baryon than anti-baryons); 3) The rapid expansion of bubble walls through the plasma induces the departure from the thermal equilibrium. All these
conditions are fulfilled by the Standard Model. However, EWBG is unable to explain the observed baryon asymmetry of the Universe
if it is only based on the Standard Model. The reason is due to the fact that EW phase transition in the Standard Model is of the first
order if the Higgs mass is constrained by $m_H \lesssim 70$GeV, in disagreement with experimental lower bound obtained
from LEP II experiment, i.e. $m_H \gtrsim 114$GeV, as well as, from recent LHC results that give a value of the Higgs mass near to $125$GeV.
Recent studies however, open the possibilities to reconsider EWBG as an available candidate for the generation of baryon asymmetry\cite{servant}.
Moreover, the EWBG mechanism is also affected by the problem related to the CP violation because the latter
generated by the Cabibbo-Kobayashi-Maskawa phase is unable to generate large enough chiral asymmetry. For a recent review on EWBG see\cite{musolf}.

  \item An interesting idea to mention is the Baryogenesis generated through evaporation of primordial BHs\cite{BHs,dolgov2}.
\end{itemize}

{\tiny
\begin{table}[]
\caption{Models of baryogenesis}
{\begin{tabular}{@{}cccc@{}} \toprule
 Mechanism &  Model  & Status/requirement  \vspace{0.09in} \\ \toprule
 Electroweak   & non-SUSY  & Excluded \\
  ($\partial_\mu J^\mu_{B/L}\sim F^2$)  &  & $m_H^{theory}< 84$GeV   \\
 ($B-L=0, B+L\neq 0$)   &   &   $m_H^{exp}>114$GeV (LEP) \\
 (triangle anomaly)   &  SUSY  & Marginal  \\
 &  & (special choice of parameters, \\
 & & e.g. $\displaystyle{\frac{m_{\text{s-top}}}{m_{\text{top}}}<1}$) \\  \hline
 Original GUT  & non-SUSY/SUSY  & Does not work (the $m_{X,Y}$  \\
 & & are to low to satisfy the\\
  & &  out equilibrium condition) \\
   & Decay of coloured  & Large baryon asymmetry \\
   & Higgs particles & can be generated\\ \hline
 GUT with Majorana interaction     & non-SUSY/SUSY  & valid if $10^{-2}$eV$< m_{\nu}<1$eV \\
   & & ($m_{\nu^M_R}>m_{\text{Higgs coloured}}$)  \\  \hline
  $\nu_R$ decay leptogenesis & non-SUSY/SUSY & valid if $m_{\nu} < 10^{-3}$eV \\
  & & ($\nu_M$ decay violates L-number)    \\ \hline 
  Affleck-Dine baryogenesis & SUSY & The model allows the so called Q-Balls \\
 & & (non topological soliton solutions).  \\
  $\varphi=(\epsilon_{\alpha\beta\gamma}{\tilde u}^c_{R\alpha}{\tilde d}^c_{R\beta}{\tilde
 d}^c_{R\gamma})^{1/3}$   &    &  Q-balls have long lifetime, and \\
 & & decay can produce huge amount of entropy \\
 $ \downarrow$  & & The problem is avoided if parameters are   \\
  (flat direction  & & chosen such that $\rho_{\text{lighest SUSY particles}}< \rho_{DM}$. \\
  of the potential) & & ${\tilde u}^c_{R}$, ${\tilde d}^c_{R}$ =\, \text{scalar partner of
  quarks} \, $u_R$, $d_R$\,, \\
  & &  $\alpha$, $\beta$, $\gamma$=\, \text{color indices}\\
  $ \uparrow$  & & \\
  Affleck-Dine leptogenesis & SUSY & $m_{\nu_1}\approx 10^{-9}$eV \\
   $\varphi_i=(2\phi_u {\tilde l}_i)^{1/2}$  & &  $\phi_u$=
  \text{Higgs field that gives $u$-quark mass}\,, \\
  & & ${\tilde l}_i$=\text{charged scalar-lepton}.\\
  & & \\ \botrule 
\end{tabular} \label{ta1}}
\end{table}
}

\section{Baryogenesis generated by coupling of baryon currents and gravitational background}
\setcounter{equation}{0}

As we have seen in the previous Section, the Baryo/Leptogenesis is generated in the framework of particle interactions (essentially GUT and SUSY).
The gravitational field enters marginally in these mechanisms. In the last years, however,
many mechanisms have been proposed in which gravity plays a fundamental role in generating the baryo/leptogenesis (see Table \ref{ta2}).
In these models matter or hadron/lepton currents are coupled
with some physical quantity characterizing the gravitational background, such as Ricci curvature or its derivative, Riemann tensor,
gravitational waves (GW),
\[
 {{\cal L}_{int} \sim J \cdot {\cal F}}\,, \quad
 J \to  \mbox{${\bar \psi} \gamma^\mu \psi, {\bar \psi} \gamma^\mu \gamma^5\psi,
 \phi \partial^\mu \phi^*,....$}\,, \quad
 {\cal F}  \to  \mbox{$R$, $\partial R$, $\partial \phi$, ...}\,.
  \]
Typically, the background is the FRW geometry, but there are also
models in which the gravitational background is described by black holes physics.

Gravitational baryogenesis share some basic features of the spontaneous spontaneous (or quintessential) baryogenesis\cite{cohen}.
In this mechanism scalar fields (or their derivatives) couple to matter or hadron/lepton current. To illustrate in some detail
the spontaneous baryogenesis, consider a neutral scalar field $\phi$. The interaction between a baryon current $J_B^\mu$ and
$\partial_\mu \phi$ is
 \begin{equation}\label{sponbaryo}
    {\cal L}=\frac{1}{M_s} J_B^\mu \partial_\mu \phi\,,
 \end{equation}
where $M_s$ characterize is a cut-off scale. In a isotropic and homogenous Universe, like FRW Universe, $\phi$ does only depend on
cosmic time, In such a case only the zero component of the baryon current ($J_B^0 = n_B$ with $n_B$ the number density of baryons)
contribute in (\ref{sponbaryo}), ${\cal L}=\mu n_B$, where $\mu \equiv {\dot \phi}/M_s$ for baryons and $\mu \equiv -{\dot \phi}/M_s$
for antibaryons. Here is assumed that the current $J_B$ is not conserved
and that, of course, ${\dot \phi}\neq 0$. The coupling (\ref{sponbaryo}) therefore gives rise to an effective chemical potential with opposite sign for
$B$ and ${\bar B}$ leading to a generation of a net baryon asymmetry even at thermal equilibrium. The latter point bypass the third Sakharov condition
because CPT violation occurs owing the Universe expansion.
The scalar field could also play the role of DE or DM. Models based on spontaneous (quintessence) baryogenesis are studied
in\cite{bento,li,liwang,feng,defelice,zhang,ahluwalia3,gu}.

\begin{table}[]
\caption{Models of gravitational baryogenesis }
{\begin{tabular}{@{}cccc@{}} \toprule
  Coupling &  $\eta\sim $  & \qquad\qquad\qquad  Comments    \\  \toprule
   & &  \\
   ${\cal L}_{int}\sim J^\mu \partial_\mu R$  & ${\dot R}$  & Gravitational Baryogenesis \cite{kitano} \\
  $J^\mu \partial_\mu R  \to E\neq {\bar E}$ & &  ${\cal L}_{int}$ can be obtained in SUGRA theories from  \\
  $\to CPT$ \text{violated}  & & higher dimensional operator or  in low energy  \\
  & & effective field theory  of quantum gravity\\ \hline
   & &  \\
  ${\cal L}_{int}\sim J^\mu\partial_\mu \phi$   & ${\dot \phi}$  &  Quintessential Baryogenesis \cite{cohen} \\
    &  &  Coupling introduced by hand \\
    & & ($\phi$ scalar field $\to$ DE/DM) \\ \hline
     & &  \\
  ${\cal L}_{int}\sim f_\mu(\phi)R{\tilde R} $  & $\frac{\mu^5}{H^{1/2}}$  &
             Leptogenesis from GWs and Inflation \cite{peskin} \\
   $R{\tilde R}= \varepsilon^{\alpha\beta\gamma\delta}
            R_{\alpha\beta\varrho\sigma}R^{\varrho\sigma}_{\quad\alpha\beta}$  & &\quad \quad  $\mu\sim 10^{14-17}$GeV  (Supersymmetric GUT)   \\
 $\partial_\mu J_l^\mu\sim R{\tilde R}$ (grav. anomalies)&
               &  CP violation $\to$ $\phi$ is a complex field (axion) as \\
               & & in SUGRA/Superstring   \\ \hline
    & &  \\
  ${\cal L}_{int}\sim h_{\mu\nu}T^{\mu\nu}\sim F_5 R{\bar\psi}\gamma^5\psi$ & ${\dot R}$
               & Gravitational Leptogenesis \cite{Lambiase:2006md} \\
   (local inertial frame)  & & {\small $T_{\mu\nu}=\bar\psi(P_f)
                                           [ F_1 P_\mu P_\nu+ F_2 \sigma_{\mu \alpha} q^\alpha P_\nu + F_3 \gamma_5  \sigma_{\mu \alpha }
           q^\alpha P_\nu$} + \\
  & & {\small $+F_4 (q_\mu q_\nu - g_{\mu \nu})+F_5 \gamma_5 (q_\mu q_\nu - g_{\mu \nu})] \psi(P_i)+(\mu \leftrightarrow \nu)$} \\
 & & ($\nabla =\partial, \Gamma=0$) \,\, $ g_{\mu\nu}=\eta_{\mu\nu}+h_{\mu\nu}$  \\
 & & {\small $g_{\mu \nu}(y)= \eta_{\mu \nu}(x) +\frac{1}{2} R_{\mu \alpha \nu \beta}(x) \, (x-y)^{\alpha}(x-y)^{\beta}+ \ldots $} \\  \botrule
\end{tabular} \label{ta2}}
\end{table}

\subsection{Gravitational Baryogenesis}

The key ingredient for the gravitational baryo/leptogenesis is a CP-violating
interaction between the derivative of the Ricci scalar curvature $R$ and the B(aryon)/L(epton) current $J^\mu$ \cite{kitano,Davoudiasl}
  \begin{equation}\label{dRJ-gravbaryo}
    {\cal L}_{int}=\frac{1}{M_*^2}\, \sqrt{-g} J^\mu \partial_\mu R
  \end{equation}
where $M_*$ is the cutoff scale of the effective theory. ${\cal L}_{int}$ is expected in a low energy effective field theory of
quantum gravity or Super gravity theories (more specifically it can be obtained in supergravity theories from a higher dimensional
operator in the K\"ahler potential).
Moreover, it {\it dynamically breaks} the CPT in an expanding Universe.
In the standard cosmological model ${\dot R}$ vanishes during the radiation era (see below). However, (tiny) deviations
from General Relativity prevent the Ricci curvature to vanish, as well as its first time derivative, so that
a net lepton asymmetry can be generated.

To generate a $B$-asymmetry, it is required that there exist $B/L$-violating processes in thermal equilibrium. In this mechanism, the
interaction $J^\mu \partial_\mu R$ gives a contribution to the energy of particles and antiparticles with opposite sign, and
thereby dynamically violates $CPT$. This coupling term modifies thermal equilibrium distribution and the chemical potential
 \begin{equation}\label{potentialsparticle}
 \mu_{particle}=\frac{\dot R}{M_*^2}=-\mu_{anti-particle}
 \end{equation}
driving the Universe towards nonzero equilibrium $B/L$-asymmetry
via the $B/L$-violating interactions. Once the temperature drops  below the decoupling temperature $T_d$
the asymmetry can no longer change and is frozen. The net asymmetry is
\begin{equation}\label{asymbaryograv}
  \eta\approx \displaystyle{\frac{\dot R}{M_*^2 T_D}}\,.
\end{equation}
In the cosmological standard model it is assumed that the energy-momentum tensor of classical fields is described by a perfect fluid
 \[
 T_{\mu\nu}=\text{diag} (\rho, -p, -p, -p)\,,
  \]
where $\rho$ is the energy density and $p$ the pressure. They are related by the relation $p=w\rho$, $w$ being the adiabatic index.
During the radiation dominated era, the equation of the state is $p=\rho/3$, i.e. $w=1/3$, and the
scale factor evolves as $a(t)=(a_0 t)^{1/2}$.
The energy density of the (classical) radiation is given by
\begin{equation}\label{rhoDM}
\rho_r=T_{00}=\displaystyle{\frac{\pi^2 g_*}{30}T^4}\,,
\end{equation}
whereas the cosmic time is related to the temperature $T$ of the Universe as
 \begin{equation}\label{t-T}
 \frac{1}{t^{2}}=\frac{32\pi^3 g_*}{90}\frac{T^4}{M_{P}^2}\,.
 \end{equation}
Moreover, Eq. (\ref{exprateH} implies that the expansion rate of the Universe can be written as
 \begin{equation}\label{bra}
   H = 1.6 g_*^{1/2} \frac{T^2}{M_P}  = -\frac{\dot R}{4R}\,.
  \end{equation}
In what follows we shall consider a flat Friedman-Robertson-Walker (FRW) Universe whose element line is
 \begin{equation}\label{metric}
    ds^2 = dt^2 -a^2(t) [dx^2+dy^2+dz^2]\,.
 \end{equation}
From the above considerations it follows that
the trace of the energy-momentum tensor of (classical) relativistic fields vanishes, $T=\rho-3p=0$.
As a consequence one has $R=-8\pi G T_\mu^\mu=0$,  and no net baryon asymmetry may be
generated. However, a possibility to generate the baryon asymmetry is given by the interaction among massless particles that
lead to running coupling constants and hence the trace anomaly \cite{kaintie}
 \[
T_\mu^\mu\propto \beta(g) F^2 \neq 0\, \qquad F^2=F^{\mu\nu}F_{\mu\nu}\,.
 \]
In a $SU(N_c)$ gauge theory with coupling $g$ and $N_f$
flavors, the effective equation of state is given by
 \[
 1-3 w=\frac{5 g^4}{96 \pi^6}  \frac{[N_c+(5/4) N_f]\,[(11/3)N_c-(2/3)N_f]}{2+(7/2)\,[N_c N_f/(N_c^2-1)]}+{\cal O}(g^5)
 \]
The numerical value of $1-3 w$ depends the gauge group
and the fermions, and  lies in the range $1-3 w \sim 10^{-2}-10^{-1}$.
The baryon asymmetry turns out to be $\eta=(1-3 w)\displaystyle{\frac{T_D^5}{M_*^2 m_P^3}}$.

Gravitational baryogenesis is conceptually similar to spontaneous baryogenesis
\cite{cohen}, see Eq. (\ref{sponbaryo}). However there some basic differences between the two paradigms:
\begin{itemize}
  \item The scalar field $\phi$ has to be added by hand, whereas the term in Eq. (\ref{dRJ-gravbaryo}) is expected
to be present in an effective theory of gravity.
  \item The scalar field $\phi$ must satisfies specific initial conditions, that is to generate a net asymmetry
   $\phi$ has to evolve homogeneously in one direction versus the other and must be
   spatially uniform. In the gravitational baryogenesis, instead, the time-evolution of $R$ naturally occurs in a cosmological
   background and it is highly spatially uniform owing to high homogeneousity of the Universe.
  \item In the regime in which $\phi$ oscillates around its minimum ${\dot \phi}$ is zero, so that the asymmetry is canceled\cite{srednicki},
    whereas the mean value of ${\dot R}$ does not vanish because is proportional to $\sim H^3$.
\end{itemize}

\subsection{Genaralised Gravitational baryogenesis}

An interesting model related to the gravitational baryogenesis has been provided by Li, Li and Zhang \cite{li},
who consider a generalized coupling of the form
 \begin{equation}\label{Jpartialf}
    {\cal L}_{int} \sim J^\mu \partial_\mu f(R)\,,
 \end{equation}
where $f(R)$ is a generic function of the scalar curvature. This function has been chosen as $f\sim \ln R$ so that
the effective interaction Lagrangian density reads
  \begin{equation}\label{Jpartialf1}
    {\cal L}_{int} \sim -c \frac{\partial_\mu R}{R} J^\mu \,,
 \end{equation}
where $c$ is a constant fixed to in order to reproduce the observed baryon asymmetry.
Following the same reasoning leading to (\ref{potentialsparticle}) one gets
 \[
 \mu_{particle}=-c \frac{\dot R}{R}=-\mu_{anti-particle}\,.
  \]
During the radiation dominated era one obtains that a net baryon asymmetry can be generated and is given by
 \begin{equation}\label{asymmZhang}
    \eta= -\frac{15g_b}{4\pi^2 g_*} \frac{c {\dot R}}{RT}\Big|_{T_D}= \frac{15}{\pi^2}\frac{cg_b H(T_D)}{g_* T_D}\simeq 0.1 c \frac{T_D}{M_P}\,,
 \end{equation}
where Eq. (\ref{exprateH}) has been used and $T_D$ is the  decoupling temperature. Moreover, one can also determine an order of magnitude of the absolute neutrino mass compatible with the current
cosmological data, i.e. $m_\nu \simeq {\cal O}(1)$eV. The idea goes along the line traced in Section \ref{sect4.1}.
In the Standard Model, $B-L$ symmetry is exactly conserved ($\partial_\mu J^\mu_{B-L}=0$). In \cite{li} the
$B-L$ violation is parameterized by higher dimensional operators, i.e. by the dimension 5 operator ${\cal L} \sim C {\bar l}l \phi^\dagger \phi$ (see Eq. (\ref{dim5})). $C$ is a scale of new physics beyond the Standard Model which generates the $B-L$ violations, $l$ and $\phi$ are the
left-handed lepton and Higgs doublets, respectively. When the Higgs field gets a vacuum expectation value $\langle \phi \rangle = v$,
the left-handed neutrino becomes massive $m_{\nu}\simeq C v$. Comparing the lepton number violating rate induced by the interaction ${\cal L}$,
$\Gamma\sim T^3$ (Eq. (\ref{Gamma1})), with the expansion rate of the Universe, $H\sim T^2$ (Eq. (\ref{exprateH})), one gets the decoupling temperature below which the lepton asymmetry is freeze-out, i.e. $T_D\simeq 10^{10}$GeV. The observed baryon asymmetry $\eta\sim 10^{-10}$ follows for $c\sim {\cal O}(1)$. Then,
assuming an approximate degenerate masses, i.e. $m_{\nu_1} \sim m_{\nu_2} \sim m_{\nu_3}$, one gets $m_\nu \lesssim {\cal 1}$eV.
The current cosmological limit comes from WMAP Collaboration \cite{spergel[17]} and SDSS Collaboration \cite{tegmark[18]}.
The analysis of Ref.\cite{spergel[17]} gives $\sum_i m_{\nu_i} < 0.69$ eV. The analysis
from SDSS gives\cite{tegmark[18]} $\sum_i m_{\nu_i} < 1.7$ eV.

\subsection{Baryogenesis in Randall-Sundrum braneworld}

The asymmetry baryon-antibaryon can arise in the Randall-Sundrum brane world model\cite{Randall}
with bulk fields owing to the effects of higher dimensionality. These studies have been performed in\cite{davaliBaryo,koyama,alberghiRan}.
The total action contains the bulk and brane actions\cite{koyama,marteensRan}
 \begin{eqnarray}\label{braneaction}
    S &=& S_{\text{bulk}}({}^{(5)}R, \Lambda, \Phi) + S_{\text{brane}}(\sigma) = \\
    &=& \int d^5 x \sqrt{G}\left[\frac{M_5}{2} {}^{(5)}R(G)-\Lambda-|\nabla_{x_M} \Phi|^2\right]+\int d^4 x \sqrt{-g}\left[\sigma+{\cal L}_{\text{matter}}\right]\,,
    \nonumber
 \end{eqnarray}
where $G_{MN}$ is the 5-dim bulk metric and $G$ its determinant, $g_{\mu\nu}$ the brane induced metric and $g$ the determinant,
$\Lambda$ the bulk cosmological constant, $\sigma$ the brane tension, and $\Phi$ the bulk complex scalar field (localized on the brane
as the graviton). $\Lambda$ and $\sigma$ are related by $\Lambda= - \displaystyle{\frac{\sigma^2}{6M_5}}$.

It is worth to write down the the effective theory on the brane. It is derived by making use of the braneworld holography\cite{koyama,hologrbrane}.
This method gives
\begin{eqnarray}\label{SeffBrane}
    S_{\text{eff}}&\simeq & \int d^4 x \sqrt{-g}\left[\frac{M_4}{2}R(g) + {\cal L}_{\text{matter}}-|\nabla_x \varphi|^2- \right.\\
    &-& \left. \frac{\log \epsilon}{4}\frac{M_4^4}{M_5^6} \left(-4R_{\mu\nu}\nabla^\mu \varphi\nabla^\nu \varphi+
     \frac{4}{3} R |\nabla \varphi|^2+R_{\mu\nu} R^{\mu\nu}-  \right. \right.\nonumber \\
    &-& \left. \left.   \frac{R^2}{3}+\frac{2}{3}|\nabla \varphi|^4 +2|(\nabla \varphi)^2|^2\right)
    +2|\nabla^2 \varphi|^2\right]+\Gamma_{\text{CFT}}\,, \nonumber
\end{eqnarray}
Here $\Gamma_{\text{CFT}}$ is the effective action for the holographic CFT on the brane, $R(g)$ is the Ricci scalar on the brane,
$M_4^2=lM_5^3=M_P^2$ plays the role of Planck mass, with $l$ the the curvature radius of the AdS spacetime, and
${\cal L}_{\text{matter}}$ is the Lagrangian density matter localized on the brane. The parameter $\epsilon$ determines the
renormalization scale of CFT, whereas the field $\phi$ corresponds to the zero mode of the bulk complex scalar field $\Phi$
localized on the brane ($\varphi$ could represent squarks or sleptons on the brane carrying baryon/lepton number).
Notice that (\ref{SeffBrane}) is written as and Hilbert-Einstein action
 \begin{equation}\label{hilbert}
    S_{EH}= \frac{1}{2\kappa^2}\int d^4x \sqrt{-g}\, R\,.
 \end{equation}
plus scalar field (in the so called Jordan frame). In this respect it is similar to scalar tensor theories.

The current associated to $\phi$, defined as $J^\mu = -i \phi \overleftrightarrow \nabla^\mu \phi^*$, satisfies the relation\footnote{Notice that
due to $U(1)$ symmetry there is a conserved current given by
 \[
 {\bar J}^\mu = J^\mu-\frac{M_4^4}{M_5^6}\left[4 R^\mu_\nu J^nu-\frac{R}{3} J^\mu-\frac{4}{3}|\nabla \varphi|^2 J^\mu+i 4
 \left(\varphi^*\nabla^\mu \varphi (\nabla \varphi^*)^2-\varphi \nabla^\mu \varphi^* (\nabla \varphi)^2\right)\right]\,.
 \]}
 \begin{equation}\label{divJphi}
    \nabla_\mu J^\mu \simeq \frac{M_4^4}{M_5^6}\left[\frac{2}{3} J^\mu \nabla_\mu R+4R_{\mu\nu}\nabla^\mu J^\nu\right] + {\cal O}(\log \epsilon)\,.
 \end{equation}
Assuming a coupling of the form (\ref{sponbaryo}), with $\phi$ a scalar field on the brane and $J_B^\mu$ replaced by $J^\mu$, one obtains (after an integration by parts) that the effective Lagrangian density on interaction is
 \begin{equation}\label{Lagrbrane}
    {\cal L} \simeq \frac{M_4^4}{M_5^6} \phi \left[\frac{2}{3}J^\mu \nabla_\mu R+4 R_{\mu\nu}\nabla^\mu J^\nu\right]\,.
 \end{equation}
This interaction leads to the baryon asymmetry given in (\ref{asymbaryograv}).
A comparison with (\ref{dRJ-gravbaryo}) suggest $M_* = fM_5^3/M_4^2$. To determine the baryon asymmetry one needs
to evaluate ${\dot R}$. In the Randall-Sundrum model, the geometrical projection method yields the field equation
 \begin{equation}\label{fieldeqbrane}
    R_{\mu\nu}-\frac{1}{2}g_{\mu\nu}R=\frac{1}{M_4^2} T_{\mu\nu}+\frac{1}{M_5^6}\pi_{\mu\nu}-E_{\mu\nu}\,,
 \end{equation}
where $T_{\mu\nu}$ is the energy-momentum tensor on the brane, $E_{\mu\nu}$ is the Weyl tensor, and
 \[
 \pi_{\mu\nu}=-\frac{1}{4}T_{\mu\alpha}T^\alpha_{\phantom{\alpha}\nu}+\frac{1}{12}T_\mu^\mu \, T_{\mu\nu}+\frac{1}{8}g_{\mu\nu}T_{\alpha\beta} T^{\alpha\beta}
 -\frac{1}{24}g_{\mu\nu} (T_\mu^\mu)^2\,.
 \]
In deriving (\ref{fieldeqbrane}) it is assumed the contribution to gravity is dominated by matter field ${\cal L}_{\text{matter}}$. Notice that the energy-momentum tensor satisfies the continuity equation $\nabla_\mu T^{\mu\nu}=0$. The trace of (\ref{fieldeqbrane})
in a FRW Universe reads
 \begin{equation}\label{Rinbrane}
 R = -\frac{T_\mu^\mu}{M_4^2}-\frac{\pi_\mu^\nu}{M_5^6}=(1-3w)\frac{\rho}{M_4^2}-(1+3w)\frac{\rho^2}{6M_5^6}\,,
 \end{equation}
from which it follows
 \begin{eqnarray}
 {\dot R} &=& -3(1+w) H \rho \left[\frac{1-3w}{M_4^2}-\frac{(1+3w)\rho}{3M_5^6}\right] \label{dotRbrane} \\
          &\simeq & \frac{8}{3}\frac{H\rho^2}{M_5^6}\sim \frac{T^{10}}{M_4 M_5^6}\,, \label{dotRbrane1}
 \end{eqnarray}
where (\ref{dotRbrane1}) follows in a Universe radiation dominated ($w=1/3$). One can compute the decoupling temperature\cite{koyama}
$T_D\sim \displaystyle{\frac{M_5^{3/2}}{M_4^{1/2}}}$, so that the bet baryon asymmetry assumes the form
 \begin{equation}\label{asymmetrybranek}
    \eta \simeq 10^{-10} \left(\frac{10^{-3}}{f}\right)^2 \left(\frac{10^8 \text{GeV}}{M_5}\right)^{12}\left(\frac{T_D}{10^{2.5}\text{GeV}}\right)^9\,,
 \end{equation}
which has been written to emphasize the estimations that the parameters characterizing the theory must assume in order that the observed baryon asymmetry
is obtained.

\vspace{0.1in}

Other models based on gravitational baryogenesis can be found in\cite{kitano,Sadjadi,Sadjadi1,sasadi,prokopec,alberghi,koyama,carroll,hamada}
\cite{Lambiase:2006dq,lambiaseConf,lambiasePLB,singh,Mohanty:2005ud,Lambiase:2006md,Lambiase:2011by,MosqueraCuesta:2009tf}.

\section{Leptogenesis by curvature coupling of heavy neutrinos}\label{Section4}
\setcounter{equation}{0}

In this Section, we study the generalization in the matter Lagrangian by including higher order terms in $R$ consistent with general covariance, Lorentz-invariance in a locally inertial frame. The effect of spin-gravity coupling will be neglected (they will be extensively discussed
in Section \ref{SectionSpin}). Therefore we work in the approximation for which the characteristic time of spinor fields variation is
smaller than the age of the Universe.

Consider the action for a four component Dirac fermion $\psi$ which couples to background gravity\cite{Lambiase:2012tn}
\begin{equation}\label{hR}
S_m[g_{\mu \nu},\psi] = \int d^4x \left[i\bar \psi \gamma^\mu (\overrightarrow\partial_\mu-\overleftarrow\partial_\mu) \psi
  - h_1(R)\, \bar \psi \psi - i h_2(R)\, \bar \psi \gamma_5 \psi \right]\,,
 \end{equation}
where $h_1(R)$ and $h_2(R)$ real valued scalar functions of the curvature,
 \begin{equation}\label{h1h2def}
 h_1(R)= M + g_1(R)\,,  \qquad h_2(R)= M^\prime + g_2( R)\,.
 \end{equation}
Here $h_1$ is a generalization of the neutrino mass term. Note that since $\bar \psi \gamma_5 \psi$ transforms as a pseudo-scalar, the $h_2$ term is odd under $CP$.
 We write the four-component fermion
 \begin{equation}
 \psi=\left(
        \begin{array}{c}
          \psi_L \\
          \psi_R \\
        \end{array}
      \right)\,.
 \end{equation}
The lagrangian in terms of the two-component fields $\psi_R$ and $\psi_L$ becomes
 \begin{eqnarray}
  {\cal L}&=&i \psi_R^\dagger\, \bar \sigma^\mu (\overrightarrow\partial_\mu-\overleftarrow\partial_\mu) \psi_R
  +i\psi_L^\dagger\,  \sigma^\mu (\overrightarrow\partial_\mu-\overleftarrow\partial_\mu)   \psi_L - \nonumber\\
  &-&\,h_1(\psi_R^\dagger \psi_L +\psi_L^\dagger \psi_R)-i\,h_2(\psi_R^\dagger \psi_L-\psi_L^\dagger \psi_R)\,,
 \label{L1}
 \end{eqnarray}
where $\sigma^\mu=(I, \sigma^i)$ and $\bar \sigma^\mu=(I,-\sigma^i)$ in terms of the Pauli matrices. The $h_2$ term can be rotated away by a chiral transformation
 \begin{equation}
 \psi_L \rightarrow e^{-i \alpha/2} \psi_L \quad  \psi_R \rightarrow e^{i \alpha/2} \psi_R\,.
\end{equation}
Keeping terms to the linear order in $\alpha$, we see that the lagrangian (\ref{L1}) changes by the amount
 \begin{eqnarray}
\delta {\cal L}&=& - \psi_R^\dagger \bar \sigma^\mu \psi_R  \partial_\mu \alpha + \psi_L^\dagger \sigma^\mu \psi_L   \partial_\mu \alpha - \nonumber\\
&-&  h_1 ( i \alpha) \left( \psi_L^\dagger \psi_R - \psi_R^\dagger \psi_L \right) -i h_2 ( i \alpha)\left( \psi_L^\dagger \psi_R + \psi_R^\dagger \psi_L \right)\,.
\label{dL1}
 \end{eqnarray}
Now we choose $\alpha=-h_2/h_1$ to eliminate the chiral mass term and obtain for the total Lagrangian
 \begin{eqnarray}
{\cal L}&=&i\psi_R^\dagger\, \bar \sigma^\mu (\overrightarrow\partial_\mu-\overleftarrow\partial_\mu) \psi_R
  +i\psi_L^\dagger\,  \sigma^\mu (\overrightarrow\partial_\mu-\overleftarrow\partial_\mu) \psi_L - \nonumber\\
  &-& \psi_L^\dagger \sigma^\mu \psi_L  \partial_\mu \left(\frac{h_2}{ h_1}\right)+ \psi_R^\dagger \bar \sigma^\mu \psi_R   \partial_\mu \left(\frac{h_2}{ h_1}\right)-\nonumber\\
  &-& \frac{1}{h_1} (h_1^2 + h_2^2) \left( \psi_L^\dagger \psi_R + \psi_R^\dagger \psi_L \right)\,.
 \label{l2a}
 \end{eqnarray}
If $h_1$ and $h_2$ are constants then, one can always rotate the axial-mass term away. We will assume that the neutrino mass $M \gg g_1$ therefore $h_1 \simeq M$ and since a constant $M^\prime$ can be rotated away $h_2= g_2$. Further we will assume that the background curvature is  only dependent on time. The lagrangian (\ref{l2a}) then reduces to the form
 \begin{eqnarray}
{\cal L}&=&i \psi_R^\dagger\, \bar \sigma^\mu (\overrightarrow\partial_\mu-\overleftarrow\partial_\mu) \psi_R
  +i \psi_L^\dagger\,  \sigma^\mu (\overrightarrow\partial_\mu-\overleftarrow\partial_\mu) \psi_L - \nonumber\\
  &-& \psi_L^\dagger \psi_L  \left(\frac{\dot g_2}{M}\right)+ \psi_R^\dagger \psi_R \left(\frac{\dot g_2}{M}\right) - \nonumber\\
  &-&  M \left( \psi_L^\dagger \psi_R + \psi_R^\dagger \psi_L \right)\,.
 \label{l2}
 \end{eqnarray}
The equation of motion for the left and the right helicity fermions derived from (\ref{l2}) are
\begin{eqnarray}
 i\bar \sigma^\mu {\partial_\mu} \psi_R
 + \left(\frac{\dot g_2}{M}\right)  \psi_R- M \psi_L =0\,,  \nonumber\\
i \sigma^\mu {\partial_\mu} \psi_L
 - \left(\frac{\dot g_2}{M}\right) \psi_L - M   \psi_R =0\,.
 \label{eom1}
 \end{eqnarray}
Written in momentum space $\psi(x)=\psi(p) e^{i (Et -\vec p \cdot \vec x)}$ the equation of motion of $\psi_R$ and $\psi_L$ are
 \begin{eqnarray}
 \left(E_R -\frac{\dot g_2}{M}\right)  \psi_R-  \vec \sigma \cdot \vec p \psi_R - M   \psi_L =0\,, \nonumber\\
 \left(E_L +\frac{\dot g_2}{M}\right)  \psi_L+  \vec \sigma \cdot \vec p \psi_L - M   \psi_R =0\,.
 \label{eom2}
 \end{eqnarray}
In the limit $p\gg M, g_2$ the dispersion relations are\cite{mukhop} $E_{R,L} \simeq E_p
\pm 2 \displaystyle{\frac{{\dot g}_2}{M}}$, where $E_p=p + \displaystyle{\frac{M^2}{2p}}$.

The canonical momenta of the $\psi_L$ and $\psi_R$ fields are as usual
 \begin{equation}
\pi_L=\frac{\partial {\cal L}}{\partial \dot \psi_L}=i \psi_L^\dagger,
\quad \pi_R=\frac{\partial {\cal L}}{\partial \dot \psi_R}=i \psi_R^\dagger\,,
\end{equation}
so that the canonical Hamiltonian density is
 \begin{eqnarray}
{\cal H}&\equiv& \pi_L \dot \psi_L +\pi_R \dot \psi_R -{\cal L}\nonumber\\
&=&i \psi_L^\dagger \dot \psi_L +i \psi_L^\dagger{\bf \sigma}\cdot {\bf \nabla}\psi_L +i \psi_R^\dagger \dot \psi_R
-i \psi_R^\dagger {\bf \sigma}\cdot {\bf \nabla}\psi_R + M  \left( \psi_L^\dagger \psi_R + \psi_R^\dagger \psi_L \right)\nonumber\\
& & +n_L \left(\frac{\dot g_2}{M}\right)-n_R \left(\frac{\dot g_2}{M}\right)\,,
\end{eqnarray}
where we have introduced the number density operators of the left and right chirality modes,
\begin{equation}
n_L\equiv  \psi_L^\dagger \psi_L , \quad n_R \equiv \psi_R^\dagger \psi_R\,.
\end{equation}
The partition function in terms of this effective Hamiltonian is
\begin{equation}
{\cal Z}= \text{Tr} e^{-\beta {\cal H}}\equiv \text{Tr} e^{-\beta( {{\cal H}_0-\mu_L n_L-n_R \mu_R})}\,,
\end{equation}
where $\beta=1/T$ and ${\cal H}_0$ is the free particle Hamiltonian. We see
that when $\dot g_2$ is non-zero then the {\it effective} chemical potential for the left chirality neutrinos is $\mu_L=-\dot g_2/M$ and for the right-chirality neutrinos is $\mu_R= \dot g_2/M$.
In the presence of interactions which change $\psi_L \leftrightarrow \psi_R$ at thermal equilibrium there will be a net difference between the left and the right chirality particles,
 \begin{eqnarray}\label{nchiralitypart}
n_R-n_L&=&\frac{1}{\pi^2}\int d^3 p \left[\frac{1}{1+e^{\beta({E_p-\mu_R})}}-\frac{1}{1+e^{\beta({E_p-\mu_L})}}\right]\\
&=& \frac{T^2}{3}\frac{\dot g_2}{M}\,. \nonumber
 \end{eqnarray}
Here we consider the simplest case in which $h_2$ and $g_2$ are linear function of the curvature $R$,
 \[
 h_2(R)=g_2(R)=\frac{R}{M_P}\,.
 \]
The axial term in (\ref{hR}) is a CP violating interaction between fermions and the Ricci curvature described by the dimension-five
operator \cite{Lambiase:2006md,Lambiase:2011by}
\begin{equation}
{\cal L}_{\diagup{\!\!\!\!\!\!C\!\!P} }= \sqrt{-g}\,\frac{1}{m_P} \, R
\bar \psi \,i\gamma_5 \psi\,. \label{cpv0}
\end{equation}
This operator is invariant under Local Lorentz transformation and
is even under $C$ and odd under $P$ and conserves $CPT$.
In a non-zero background $R$, there
is an effective $CPT$ violation for the fermions. Take $\psi=(N_R, N_R^c)^T$, where $N_R$ is a heavy right handed neutrino and $N_R^c$ a left handed heavy neutrino, which decay into the light neutrinos. Majorana neutrino interactions with the light neutrinos and Higgs  relevant for leptogenesis, are
described by the lagrangian
 \begin{equation}
 {\cal{L}}=- h_{\alpha \beta}(\tilde{\phi^\dagger}~
 \overline{N_{R \alpha}} l_{L \beta})
-\frac{1}{2}N_R^c \,{\widetilde M}\,N_R +h.c. \,,
\label{LN}
 \end{equation}
where ${\widetilde M}$ is the right handed neutrino mass-matrix, $l_{L \alpha}=
(\nu_\alpha , e^-_\alpha)_L^T $ is the left-handed lepton doublet
($\alpha $ denotes the generation), $\phi=(\phi^+,\phi^0)^T $ is
the Higgs doublet. In the scenario of leptogenesis introduced by
Fukugita and Yanagida, lepton number violation is
achieved by the decays $N_R \rightarrow \phi + l_L$ and also
${N_R}^c \rightarrow \phi^\dagger + {l_L}^c$. The difference in
the production rate of $l_L$ compared to $l_L^c$, which is
necessary for leptogenesis, is achieved via the $CP$ violation. In
the standard scenario, $n(N_R)=n(N_R^c)$ as demanded by $CPT$, but
 \[
 \Gamma(N_R \rightarrow l_L + \phi) \not = \Gamma(N_R^c \rightarrow l_L^c + \phi^\dagger)
 \]
due to the complex phases of the Yukawa coupling matrix $h_{\alpha \beta}$, and a net lepton
number arises from the interference terms of the tree-level and
one loop diagrams (see Section 2 and Ref.\cite{luty,Flanz}).

In this leptogenesis scenario we have that the decay rates of $N_R$ and $N_R^c$ are the same,
 \[
 \Gamma(N_R \rightarrow l_L + \phi) = \Gamma(N_R^c \rightarrow l_L^c + \phi^\dagger)\,,
  \]
but there is a difference between the heavy light and left chirality neutrinos at thermal equilibrium due to the CP violating gravitational interaction (\ref{cpv0}),
\begin{equation}
 n(N_R)-n(N_R^c)=\frac{T^2}{3} \frac{\dot R}{M_P M}\,.
\label{delta N}
\end{equation}
The $N_R \leftrightarrow N_R^c$ interaction can be achieved by the scattering with a Higgs field.
A recent example of leptogenesis due  heavy neutrino decay with CP violation in a SO(10) model is described in\cite{Abada}.
In standard SO(10) unification, all Standard Model fermions of a given generation together with a right-handed neutrino are in a ${\bf 16}$ representation of SO(10),
 \begin{eqnarray}
{\bf 16_f} &=& ({\bf 1_f} + {\bf \bar 5_f} + {\bf \bar{ 10}_f})_{SU(5)} \nonumber\\
&=& (N_R + ( L, d^c)+ (Q, u^c, e^c))
 \end{eqnarray}
The charged fermion and Dirac neutrino mass matrices receive contributions from Yukawa couplings of the
form $\bf{16_f 16_f H}$ (where ${\bf H = 10_H, 126_H}$ and/or ${\bf 120_H}$).  Majorana masses for the right-handed neutrinos are generated either from
\begin{equation}
{\bf 16_f\, 16_f \,{\overline {126}}_H} \supset y \,S^\prime\, N_R^c \,N_R
\label{MNr1}
\end{equation}
or from the non-renormalizable operators suppressed by some mass scale $\Lambda$
\begin{equation}
 \frac{f}{\Lambda}{\bf 16_f\, 16_f \,{\overline{ 16}}_H \,{ \overline{ 16}}_H} \supset \frac{f}{\Lambda}\, S^2\, N_R^c\, N_R\,.
 \label{MNr2}
\end{equation}
When the GUT Higgs fields $S^\prime$ or $S$ acquire a {\it vev}, a large Majorana  mass $M$ is generated for $N_R$ which breaks lepton number spontaneously. This following the see-saw mechanism leads to small neutrino masses at low energies. At temperatures larger than the heavy neutrinos and the GUT Higgs masses one there will be helicity flip scattering interactions like $S + N_R \leftrightarrow S+  N_R^c$ which change the lepton number (as $T> M$ the helicity and the chirality of $N_R$ are same). The interaction rate is
\begin{equation}
\Gamma(S N_R\leftrightarrow S N_R^c)= \langle n_s \sigma \rangle= \frac{0.12}{\pi} \left(\frac{f}{\Lambda}\right)^2 T^3\,.
\end{equation}
The interactions decouple at a temperature $T_D$. The latter is computed via the equality
\begin{equation}\label{H=T}
 \Gamma(T_D)=H(T_d)\,,
\end{equation}
where $H={\dot a}/a$.
From (\ref{H=T}) one derives the decoupling temperature
\begin{equation}
T_D=13.7 \pi \sqrt{g_*} \left(\frac{\Lambda}{f} \right)^2 \frac{1}{M_P}=
  13.7 \pi \sqrt{g_*} \left(\frac{\langle S\rangle^2}{M} \right)^2 \frac{1}{M_P}\,,
\label{TD}
\end{equation}
where we have used $M=f \langle S \rangle/\Lambda$.

From the lepton asymmetry (\ref{delta N}) and (\ref{sentropy}) one obtains the value of frozen in lepton
asymmetry as
\begin{equation}\label{eta}
\eta= \frac{n(N_R)-n(N_R^c)}{s}= \frac{15}{2 \pi^2 g_*}\frac{\dot R(T_D)}{T_D \,M \,M_P }\,,
\end{equation}
This result agrees with\cite{kitano}.

Some comments are in order. $1)$ In the case in which the fermion is, for example, an electron one also gets a splitting of energy levels
$E(e_R ) - E(e_L)$, but this does not lead to lepton generation of lepton asymmetry as both $e_L$ and $e_R$ carry the same lepton number.
$2)$ In principle, one should also take into account primordial perturbations of the gravitational background (characterized mainly by scalar and tensor perturbations) and of the energy density and pressure, characterized by $\delta \rho=\delta T_0^0$ and $\delta p \delta_i^j=\delta T_i^j$ (see for example \cite{ma}). These perturbations are related as $\delta p = c_s^2 \delta \rho$, where $c_s^2=w+\rho dw/d\rho$ is the adiabatic sound speed squared. For relativistic particles  $w=1/3$ and therefore $c_s^2=1/3$.
As a consequence, the trace of the perturbed energy-momentum tensor vanishes (this is not true
in presence of anisotropic shear perturbations), so that according to the gravitational leptogenesis mechanism, no net baryon asymmetry can be generated.

\subsection{Avoiding subsequent wash-out}\label{sect4.1}

The light neutrino asymmetry can be erased by the interactions $\nu_L +\phi_0 \rightarrow \nu_L^c +\phi_0^\dagger$ with the
standard model Higgs. To prevent the erasure of the lepton asymmetry by Higgs scattering, we must demand that the lightest
heavy neutrino mass be lower than the decoupling temperature of
the light-neutrino Higgs interaction, which is calculated as
follows. The light neutrino masses arise from an effective
dimension five operator (\ref{dim5}) which is obtained from
(\ref{LN}) by heavy neutrino exchange\cite{Weinberg}
\begin{eqnarray} \label{dim5}
 {\cal{L}} &=&  C_{\alpha \beta} \,(\overline{ {l_{L \alpha}}^c } ~ \tilde{\phi^*})(\tilde{\phi^\dagger}~ l_{L \beta}) +  h.c. \\
 &=&  \frac{C_{\alpha \beta} }{2M} (\overline {{l_{L \alpha a}}^c} ~\epsilon^{a m} \phi_m) (l_{L \beta b}
~\epsilon^{bn} \phi_n)
 + \frac{C_{\alpha \beta}^* }{2M} (\overline {l_{L \alpha a}} ~\epsilon^{a m} \phi_m^*)
  ({l_{L \beta b}}^c ~\epsilon^{bn} \phi_n^*)\,. \nonumber
 \end{eqnarray}
Here  $\tilde \phi\equiv i \sigma_2 \phi^*=(-{\phi^0}^*, \phi^-)^T$ , $\epsilon^{a b}$ is the antisymmetric tensor, and $a,b..$ denote the
gauge $SU(2)_L$ indices.

The $\Delta L=2$ interactions that result from the operator (\ref{dim5}) are
 \begin{eqnarray}
 \nu_L + \phi^0 \longleftrightarrow \nu_R + \phi^0\,,
 \nonumber \\
 \nu_R + {\phi^0}^*
 \longleftrightarrow \nu_L + {\phi^0}^*\,.
 \label{ints}
 \end{eqnarray}
The cross section for the interaction $\nu_{L \alpha} +\phi^0 \leftrightarrow
\nu_{R \beta} +\phi^0$ is
\begin{equation}
 \sigma=\frac{|C_{\alpha \beta}|^2}{2 M^2} \frac{1}{\pi}\,,\label{cs}
 \end{equation}
In the electroweak era, when the Higgs field in (\ref{dim5})
acquires a $vev$, $\langle\phi^0\rangle =v=174~ GeV$, this
operator gives rise to a Majorana neutrino mass matrix
 \[
m_{\alpha \beta}=4 v^2 \, \frac{C_{\alpha \beta}}{M}\,,
 \]
and the cross section (\ref{cs}) can be expressed in terms of light neutrino masses as
 \begin{equation}\label{cs2}
 \sigma=\frac{|m_{\alpha \beta}|^2}{32 \pi v^4}\,.
  \end{equation}
The interaction rate of the lepton number violating scattering
$\nu_{L} +\phi_0 \leftrightarrow \nu_{R} +\phi_0^\dagger$
is given by
 \[
 \Gamma(\nu_{L} +\phi_0 \leftrightarrow \nu_{R} +\phi_0^\dagger) = \frac{0.122}{16 \pi} \frac{m_\nu^2 \, T^3}{v^4}\,.
 \]
The decoupling temperature $T_{l}$ when the interaction rate
$\Gamma (T_l)$ falls below the expansion rate of the Universe (\ref{exprateH}.
The decoupling temperature $T_l$ is obtained from equation $\Gamma (T_l)=H(T_l)$, where
 \begin{equation}\label{Gamma1}
 \Gamma (T_l)=
 \frac{0.122}{\pi}~\frac{|C_{\alpha \beta}|^2 T_l^3}{M^2} = 1.7 \sqrt{g_*} ~\frac{T_l^2}{M_P}\,.
 \end{equation}
It turns out that
\begin{equation}\label{Tl}
 T_l= 2 \times 10^{14}\,  \left(\frac{0.05 eV}{m_{\nu}} \right)^2 \text{GeV}\,,
 \end{equation}
The heavy neutrino decays occur at $T\simeq M \simeq 10^{12}$GeV, below the temperature $T_l\simeq 2 \times 10^{14}$GeV.
At temperatures $T\sim T_l$ the light-neutrino lepton number violating interactions are effective. As a consequence,
the lepton number asymmetry from the decay of asymmetric number of heavy neutrino decays is not washed out by Higgs
scattering with light neutrinos.

\section{Models and Time varying Ricci curvature in different cosmological scenarios}
\setcounter{equation}{0}

We  now discuss some cosmological scenarios in which the gravitational leptogenesis mechanism
can be realized.

\subsection{Gravitational Leptogenesis in $f(R)$ theories of gravity}

As discussed in the Introduction, the observation that the present phase of the expanding Universe is accelerated
has motivated in the last years the developments of many models of gravity which go beyond the general relativity, and therefore the standard
cosmological model. Among the different approaches, the $f(R)$-theories of gravity have received a great attention.
The reason relies on the fact that they allow to explain, via a gravitational dynamics, the
observed accelerating phase of the Universe, without invoking exotic matter as sources of dark energy.
Moreover, they also provide an alternative approach to expliain Dark Matter problem.

The Lagrangian density of these  models does depend on higher-order curvature
invariants(see\cite{silvestri,frieman,maartens,faraonibook,defeliceLiving,cliftonPhysRep,NojiriPhysRep} and references therein),
such as, for example, $R^2$, $R_{\mu\nu}R^{\mu\nu}$, $R\Box R$, and so on.
Here we focalize our attention to $f(R)$ models which are a generic function of the Ricci scalar curvature $R$
\begin{equation}\label{Lagr}
  S=\frac{1}{2\kappa^2}\int d^4x \sqrt{-g}\, f(R)+S_m[g_{\mu\nu},\psi]\,.
\end{equation}
In (\ref{Lagr}), $S_m$ is the action of matter and $\kappa^2=8\pi G=8\pi M_P^{-2}$ ($M_P\simeq 10^{19}$GeV is the Planck mass).
Cosmological and astrophysics consequences of (\ref{Lagr}) have been largely studied in literature
\cite{nojiri,capozziello-f,delaur,mota,mishra,capaldo,Capozziello:1999xt,troisi,Capozziello:1999uwa,capozziello3,capozziello1,lambiase2,
tsujikawa,appleby,amendola,bean,nojiri2008,lima,olmo1,pogosian,hu,libarroew,clifton}.

$f(R)$ gravity provide scenarios that make these models very attractive. In fact \cite{nojiri}:
1) They allow to unify the early-time Inflation and the later-time acceleration of the Universe owing to
the different role of the gravitational terms relevant at small and large scales;
2) DM and DE issues can be treated in a unique and unified setting; 3)
They provide a framework for the explanation of hierarchy problem and unification of GUT with gravity.
However, solar system tests strongly constraint or rule out many $f(R)$ models of gravity.
Therefore the form of the generic function $f(R)$ must be properly constructed.
In this respect, available models are:

\begin{itemize}
  \item The Hu and Sawicki model\cite{hu}
    \begin{equation}\label{hu}
    f(R)=-m^2\frac{c_1(R^2/m^2)^{2n}}{1+c_2(R/m^2)^{2n}}\,,
 \end{equation}
  \item The Starobinsky model\cite{starob}
   \begin{equation}\label{starob}
    f(R)=R+\lambda R_{st}\left[\left(1+\frac{R^2}{R^2_{st}}\right)^{-d}-1\right]-\alpha R^2\,,
  \end{equation}
   \item The Nojiri and Odintsov model\cite{NOmodel}
     \begin{equation}\label{NOmodelf}
    f(R)=R+\frac{\alpha R^l-\beta R^m}{1+\gamma R^n}\,.
  \end{equation}
 \end{itemize}
The parameters $c_1$, $c_2$, $d$, $m$, $n$, $l$, $\lambda$, $R_{st}$, $\alpha$, $\beta$ and $\gamma$
entering the above equations are free. Their combinations allow to get a description of cosmic acceleration
(early and present) of the Universe .

A characteristic of the models (\ref{hu})-(\ref{NOmodelf}) is that the $R$-terms can be expanded in the appropriate regimes,
reproducing simplest form of $f(R)$. A particular subclass is of the form
 \begin{equation}\label{f(R)=R+aR**n}
    f(R)=R+\alpha R^n\,,
 \end{equation}
where $\alpha>0$ has the dimensions [energy]$^{-2(n-1)}$ and $n>0$.
Particularly interesting is the case $n=2$ (referred in literature as Starobinsky's model \cite{starobinsky1980})
 \begin{equation}\label{f(R)=R+aR**2}
    f(R)=R+\alpha R^2\,.
 \end{equation}
This model (\ref{f(R)=R+aR**2}) has been studied in the framework
of astrophysics and cosmology. For instance, gravitational radiation emitted by isolated system constraints the free parameter
to $|\alpha| \lesssim (10^{17} - 10^{18})$m$^2$ \cite{berry,jetzer}. E{\"o}t-Wash experiments lead instead to the constraints
\begin{equation}\label{eot-washed}
    |\alpha|\lesssim 2 \times 10^{-9} \mbox{m}^2\,.
\end{equation}
More stringent constraints are provided by the Cosmic Microwave
Background (CMB) physics. The amplitude of the curvature perturbation corresponding to (\ref{f(R)=R+aR**2}) is $P_{\cal R}\simeq \displaystyle{\frac{N_k^2}{18\pi}\frac{1}{\alpha m_P^2}}$, with $N_k\sim 55$. Using the WMAP 5-years data\cite{WMAP-5} ($P_{\cal R}\sim 2.445 \times 10^{-9}$), it follows that $\alpha$ is constrained as \cite{defeliceLiving}
 \begin{equation}\label{WMAP}
 |\alpha| < 10^{-39}{\mbox m}^2\,.
 \end{equation}
The bound (\ref{WMAP}) is obtained in the regime $R\gg \alpha^{-1}$ (in this regime the model describes the inflationary epoch).

In these models of $f(R)$ gravity is implicitly assumed that the chameleon effect\cite{chameleon} holds, which means that
the Compton length $\lambda$ associated to the characteristic scales, coming out from adding (pertubative) higher order terms to
the Hilbert-Eisntein action, are smaller or larger in regions with higher or lower matter density. Typically one assumes that
$\lambda$ is constant, so that the theory is viewed as a local effective theory which is valid for a certain range of parameters.

\subsubsection{Field equations in $f(R)$ gravity}

The field equations obtained by the variation of the action (\ref{Lagr}) with respect to the
metric are
\begin{equation}\label{fieldeqs}
  f' R_{\mu\nu}-\frac{f}{2}\, g_{\mu\nu}-\nabla_\mu \nabla_\nu f'
  +g_{\mu\nu}\Box f'=\kappa^2 T_{\mu\nu}\,,
\end{equation}
where the prime stands for the derivative with respect to $R$. The trace reads
\begin{equation}\label{tracef}
  3\Box f'+f' R-2f=\kappa^2 T_\mu^\mu\,,
\end{equation}
In the spatially flat FRW Universe, Eq. (\ref{metric}),
Eqs. (\ref{fieldeqs}) and (\ref{tracef}) become
\begin{eqnarray}
3f' H^2-\frac{R f'-f}{2}+3H f'' {\dot R}=\kappa^2 \rho\,, \label{0-0} \\
-2 f' H^2-f''' {\dot R}^2+f''(H f'' {\dot R}-{\ddot R})=\kappa^2 (\rho +p)\,, \label{i-j} \\
3f''' {\dot R}^2+3f'' {\ddot R}+9H f'' {\dot
R}+f'R-2f=\kappa^2 T\,, \label{trace}
\end{eqnarray}
Moreover, the Bianchi identities give a further condition on the
conservation of the energy
\begin{equation}\label{EnCons}
  {\dot \rho}+3\frac{\dot a}{a}(\rho +p)=0\,.
\end{equation}
In what follows, we shall look for those solutions of field equations such that the scale factor evolves
as
 \begin{equation}\label{a(t)}
a(t) = a_0 t^\beta\,, \qquad H=\frac{\beta}{t}\,.
 \end{equation}
The scalar curvature turns out to be
 \begin{equation}\label{R}
R=6 (2H^2+{\dot H}) = {\frac{6\beta(2\beta-1)}{t^2}}\,.
 \end{equation}
The $f(R)$ model we concern here is that one of Eq.  (\ref{f(R)=R+aR**n}).
By using Eqs. (\ref{0-0}) and (\ref{i-j}) and the usual expression relating the energy density and the pressure,
$p=w\rho$, where $w$ is the adiabatic index, one gets
 \begin{equation}\label{w-n}
    w=\frac{1}{3}+\varsigma(t)\,, \quad \varsigma\equiv \frac{2}{3\beta}\left(\frac{\beta+n {\cal A}}{\beta+{\cal A}}-2\beta\right)\ll 1\,,
 \end{equation}
 with
 \[
 {\cal A}\equiv \alpha R^{n-1}[\beta(2-n)-(n-1)(2n-1)]\,.
 \]
The energy density $\rho$ assumes the form
 \begin{equation}\label{rho-n}
   \kappa^2 \rho=\frac{3\beta^2}{t^2}\left(1+\frac{\cal A}{\beta}\right)\,.
 \end{equation}
Notice that during the radiation dominated era ($\beta=1/2$), to which we are mainly interested, the quantity ${\cal A}$ vanishes
because $R=0$, as well as the perturbation $\varsigma$, and the adiabatic index reduces to the standard value $w=1/3$.
Moreover, our concern is for the regime $\alpha R^{n-1} \lesssim 1$.

\subsubsection{Constraints from BBN}

In BBN one has to consider the weak interaction rate of particles ($p, n, e^{\pm}$ and $\nu$) in thermal equilibrium. For $T\gg {\cal Q}$ (${\cal Q}=m_n-m_p$, where $m_{n,p}$ are the neutron and proton masses), one gets\cite{bernstein,kolb,Copi,burles}
$\Lambda(T)\simeq q T^5$, where $q=9.6 \times 10^{-46}\mbox{eV}^{-4}$.

The primordial mass fraction of ${}^4 He$ is estimated by defining $Y_p\equiv \lambda \, \frac{2 x(t_f)}{1+x(t_f)}$,
where $\lambda=e^{-(t_n-t_f)/\tau}$. $t_f$ and $t_n$ are the time of the freeze-out of the weak interactions and of the nucleosynthesis,
respectively, $\tau\simeq 887$sec is the neutron mean life, and $x(t_f)=e^{-{\cal Q}/T(t_f)}$ is the neutron to proton equilibrium ratio.
The function $\lambda(t_f)$ represents the fraction of neutrons that decay into protons in the time $t\in [t_f, t_n]$.
Deviations from $Y_p$ (generated by the variation of the freezing temperature $T_f$) are given by\cite{torres,Lambiase:2005kb,Lambiase:2012fv}
$\delta Y_p=Y_p\left[\left(1-\frac{Y_p}{2\lambda}\right)\ln\left(\frac{2\lambda}{Y_p}-1\right)-\frac{2t_f}{\tau}\right]
\frac{\delta T_f}{T_f}$. In the above equation we have set $\delta T(t_n)=0$ because $T_n$ is fixed by the deuterium binding energy.
The current estimation on\cite{coc} $Y_p$, $Y_p=0.2476\pm \delta Y_p$, with $|\delta Y_p| < 10^{-4}$, leads to
 \begin{equation}\label{deltaT/Tbound}
    \left|\frac{\delta T_f}{T_f}\right| < 4.7 \times 10^{-4}\,.
 \end{equation}
The freeze-out temperature $T$ is determined by $\Lambda= H$. One gets $T=T_f(1+\frac{\delta T_f}{T_f})$, where $T_f\sim 0.6$ MeV and
 \begin{equation}\label{deltaT/T}
    \frac{\delta T_f}{T_f} = \varsigma \frac{4\pi}{15}\sqrt{\frac{\pi g_*}{5}}\frac{1}{qm_P T_f^3}\simeq 1.0024 \left(\beta-\frac{1}{2}\right)\,.
 \end{equation}
Equations (\ref{deltaT/T}) and (\ref{deltaT/Tbound}) implies (see also Ref.\cite{barrow,kang})
 \begin{equation}\label{boubdc}
    2\beta-1 \lesssim 9.4 \times 10^{-4}\,.
 \end{equation}

\subsubsection{Gravitational leptogenesis induced by $f(R)$ gravity}

Using the definition of Ricci scalar curvature (\ref{R}), it follows
\begin{equation}
 {\dot R}=-\frac{12\beta(2\beta-1)}{t^3}\,.
\end{equation}
Equation (\ref{t-T}) then implies (to leading order in $(2\beta-1)$)
\begin{eqnarray}
\eta &=& \frac{128 \pi^2}{3 \sqrt{5}} \beta (2\beta-1) \sqrt{\pi g_*}
\frac{T_D^5}{M_P^4 M}\simeq  \label{eta-n}
\end{eqnarray}
 \[
 \simeq (2\beta-1) 3.4 \times 10^{-10}\, \frac{10^{12}\mbox{GeV}}{M}\left(\frac{T_D}{10^{15}\mbox{GeV}}\right)^5\,.
 \]
An inspection of (\ref{eta-n}) immediately revels that the observed baryon asymmetry can be obtained, for example,
for $T_D\sim 10^{16}$GeV, $M\sim 10^{12}$GeV (see (\ref{M1lowebouond}) and for example\cite{strumia}), provided that $2\beta-1\simeq 2\times 10^{-6}$.
The value of the heavy neutrino mass $M\sim 10^{12}$GeV is consistent with the atmospheric neutrino scale $m_\nu=0.05$ eV,
obtained from the see-saw relation $m_\nu=m_D^2/M$ with the Dirac mass scale $m_D\sim {\cal O}(10)$ GeV.

The lepton asymmetry generated via (\ref{eta-n}) is passed on to the light neutrino sector when the heavy neutrino decays at temperature $T\sim M\sim 10^{12}$GeV. The effects of washed out are avoided by
considering the effective (five dimensional) operator violating the lepton number $\Delta L=2$, as before discussed.
Notice that the baryon asymmetry is generated both for\cite{Lambiase:2012tn} $n\neq 2$ and $n=2$
The case $n < 0$ is excluded because these $f(R)$ models of gravity are affected by instability problems\cite{faraoniRMP,faraonibook,defeliceLiving}.

\subsection{Time varying Ricci curvature from quantum fluctuations}

In this Section we discuss another interesting cosmological scenario in which a non-zero Ricci curvature is generated in the radiation era by back-reaction of quantum fields. Quantum effects cannot be ignored because they may modify the dynamics of the Universe evolution. In order to incorporate these back-reaction effects in the cosmic evolution of the Universe, General Relativity requires some modification.
Again without a complete theory of quantum gravity, one works assuming a semiclassical theory of gravity\cite{birrell}.
In this context, the Einstein field equations are rewritten as\cite{birrell,anderson}
 \begin{equation}\label{Einsteineqs}
 R_{\mu\nu}-\frac{1}{2}g_{\mu\nu}R=\frac{8\pi}{M_P^2}\left(T^{(cl)}_{\mu\nu}+\langle T^{(QM)}_{\mu\nu}\rangle\right)
 \end{equation}
where $T^{(cl)}_{\mu\nu}$ is the stress energy-momentum tensor for the classical field, $T_{\mu\nu}^{(QM)}$ represents the energy momentum tensor operator generated by quantum fields, and finally $\langle T_{\mu\nu}^{(QM)}\rangle=\langle 0 | T_{\mu\nu}^{(QM)}|0 \rangle$ represents the
regularized expectation value of $T_{\mu\nu}^{(QM)}$. During the radiation dominated era, although the trace of the classical energy momentum tensor
vanishes, $T^{(cl)}=0$, the presence of the quantum corrections $\langle 0 | T_{\mu\nu}^{(QM)}|0 \rangle$
implies that the trace is nonvanishing, and therefore a net baryon asymmetry could be generated by having ${\dot R}\neq 0$.
This trace anomaly comes from the infinite counterterms that must be add to the gravitational action to make the trace finite.

The dynamical evolution of the gravitational background is assumed to be described by the FRW Universe, Eq. (\ref{metric}).
The regularized components of the energy-momentum tensor  have the form \cite{birrell,opher}
 \begin{equation}\label{Tmunu}
    \langle T_{\mu\nu}^{(QM)} \rangle= k_1\,\,{}^{(1)}H_{\mu\nu}+k_3\,\, {}^{(3)}H_{\mu\nu}\,,
 \end{equation}
where
 \begin{eqnarray}
   {}^{(1)}H_{\mu\nu} &=& 2 R_{;\mu;\nu}- 2g_{\mu\nu}  \Box R +2 R R_{\mu\nu}-\frac{R^2}{2} g_{\mu\nu}\,, \label{H}\\
   {}^{(3)}H_{\mu\nu} &=&  R_{\mu}^{\,\,\alpha}R_{\nu \alpha}-\frac{2}{3}R R_{\mu\nu}-\frac{1}{2}R^{\alpha\beta}R_{\alpha\beta}g_{\mu\nu}+\frac{R^2}{4}g_{\mu\nu}\,,
   \nonumber
 \end{eqnarray}
$\Box=\nabla_\mu \nabla^\mu$, and $;$ stands for covariant derivative.
The coefficients $k_{1, 3}$ are constants and come from the regularization process. Their values strictly depend not only on number and types of fields present in the Universe, but also on the method of regularization. Because the methods of regularization affect the the values of $k_{1, 3}$ and more important because of the uncertainty of what fields were present in the very early Universe, they can be considered as free parameters\cite{anderson,opher}.
The tensor ${}^{(1)}H_{\mu\nu}$ satisfies $\nabla_\mu {}^{(1)}H^{\mu}_{\nu}=0$. It is obtained by varying the local action
 \[
 {}^{(1)}H_{\mu\nu}=\displaystyle{2}\sqrt{-g}\frac{\delta}{\delta g_{\mu\nu}}\int d^4 \sqrt{-g} R^2\,.
 \]
The infinities in $\langle T^{(QM)}\rangle$ are canceled by adding infinite counterterms in the Lagrangian density that describes the gravitational fields. One of these counterterms if of the form $\sqrt{-g}C R^2$, and due to (the logarithmically divergent) constant $C$, the coefficients $k_1$ is arbitrary(actually it can be fixed experimentally\footnote{It is worth nothing that by making use of the dimensional regularization \cite{duff}, for example, one infers
 \[ k_1=\frac{1}{1440\pi^2}\left(\frac{1}{2}N_0+3N_{1/2}+6N_1\right)\,. \]}).
 As regards ${}^{(3)}H_{\mu\nu}$, it is covariantly conserved only for conformal flat spacetimes, and cannot be derived by means of the variation of a local action, as for ${}^{(1)}H_{\mu\nu}$. The coefficient $k_3$ is given by
 \[
 k_3 = \frac{1}{1440\pi^2}\left(N_0+\frac{11}{2}N_{1/2}+31N_1\right)\,.
 \]
For a $SU(5)$ model, for example, the number of quantum fields take the values $N_0=34$, $N_{1/2}=45$, and $N_{1}=24$, so that\cite{opher} $k_3 \simeq 0.07$.

The explicit expression of the components of ${}^{(1)}H_{\mu\nu}$ and ${}^{(3)}H_{\mu\nu}$ are
 \begin{eqnarray}\label{Hcomponents}
    {}^{(1)}H_{00} & = & 18(2{\ddot H}H+{\dot H}^2+10 {\dot H} H^2)\,, \\
     {}^{(1)}H_{ij} &=& 6\left(2\frac{d^3H}{dt^3}+12{\ddot H} H+14 {\dot H} H^2+7{\dot H}^2\right)g_{ij}\,,
     \nonumber
 \end{eqnarray}
 \begin{equation}\nonumber
    {}^{(3)}H_{00}=3H^4\,, \quad {}^{(3)}H_{00}= H^2(4{\dot H}+3H^2)g_{ij}\,.
 \end{equation}
Applying the regularization procedure one infers the trace
anomaly\footnote{The anomaly trace is typically expressed in term of curvature tensors and their covariant derivatives, as well as mass-terms\cite{birrell,dappiaggi}, i.e.
 \[
 \langle T_\mu^\mu \rangle = \alpha(N_0, N_{1/2}, N_1)\left(R^{\mu\nu}R_{\mu\nu}-\frac{R^2}{3}\right)+c_1(m) R+
 c_2(m)+c_3\Box R-\sum_{i=1}^{N_0} m_i^2\langle \phi_i^2\rangle  -\sum_{i=1}^{N_{1/2}} m_i\langle {\bar \psi}_i \psi_i\rangle\,,
 \]
where the coefficients $c_{1, 2}(m)$ are combinations of (power) mass fields, $N_{0, 1/2, 1}$ is the number of the quantum matter of boson, fermion and vector fields. The coefficients $c_i$ are subject to a finite renormalization, becoming free parameters of the theory. For our purpose we can neglect the mass-terms since the fields are relativistic, keeping in mind however, that the procedure of renormalization gives rise to purely geometric terms which appear in the final expression of the trace anomaly.}
 \begin{equation}\label{Trace}
    \langle T^{(QM)\mu}_{\phantom{(QM)}\mu} \rangle = k_3\left(\frac{R^2}{3}-R_{\alpha\beta}R^{\alpha \beta}\right)-6k_1 \Box R\,,
   \end{equation}
that for the FRW Universe assumes the form
  \[
    \langle T^{(QM)}  \rangle = 36 k_1 \left(\frac{d^3H}{dt^3}+7\ddot{H}H+4{\dot{H}}^2+12\dot{H}H^2\right)+
    \]
 \begin{equation}\label{TraceFRW}
 +12k_3H^2\left(\dot{H}+H^2\right)\,.
  \end{equation}
During the radiation dominated era, as we have seen, the trace of the energy-momentum tensor of classical fields vanishes $T^{(cl)}=\rho-3p=0$.
The trace anomaly  instead gives a non vanishing contribution.
To evaluate it, we point out that both the scale factor $a(t)=(a_0 t)^{1/2}$  and the relation between the cosmic time and
the temperature $T$, Eq. (\ref{t-T}), should be modified by the back-reaction effects induced by quantum fields. As we shall see below
the evolution of the Universe can be described by standard cosmology. In a FRW Universe, the modified Einstein field equations assume the form
 \begin{eqnarray}\label{EisnteinEqsExplicit1}
    3H^2  &=&  \frac{8\pi}{M_P^2}\left[\rho+18k_1(2{\ddot H}H+{\dot H}^2+10 {\dot H}H^4)+3k_3 H^4\right]\,, \\
    3H^2+2{\dot H} &=& \frac{8\pi}{M_P^2}\Big[-p+6k_1\Big(2\frac{d^3H}{dt^3}+ 12{\ddot H}H+14 {\dot H}H^2+7{\dot H}^2\Big)+
                  \label{EisnteinEqsExplicit2} \\
     & &  + k_3 H^2(4{\dot H}+3H^2\Big)\Big]\,,
 \end{eqnarray}
from which it follows
 \[
    2H^2+{\dot H}=\frac{8\pi}{M_P^2}\Big[6k_1(\frac{d^3H}{dt^3}+7 {\ddot H}H +4{\dot H}^2+12{\dot H}H^2)
 \]
 \begin{equation}\label{EqH}
    +2k_3 H^2({\dot H}+H^2)\Big]\,.
 \end{equation}
We are looking for solutions of the form
 \begin{equation}\label{H=H0+delta}
    H(t)=H_0(t)+\delta(t)\,,
 \end{equation}
where $\delta(t)\ll 1$ is a perturbation, and $H_0=1/2t$ is the Hubble parameter for a  radiation dominated universe. The Ricci curvature vanishes, $R=0$, as well as its covariant and (cosmic) time derivatives. It then follows that ${}^{(1)}H_{\mu\nu}(H=H_0)=0$ whereas
${}^{(3)}H_{\mu\nu}\neq 0$ when $H=H_0$ (see\cite{Lambiase:2011by} for details). Inserting $H$ given in (\ref{H=H0+delta})
into Eq. (\ref{EqH}), one obtains (to leading order) the solution for $\delta$:
 \begin{equation}\label{Solution}
  \delta(t)\simeq \frac{k_3}{M_P^2}\frac{1}{t^3}-\frac{C}{4M_P^2}\frac{1}{t^4}\,.
 \end{equation}
$C$ is a constant of integration. As it can be seen, the $M_P^{-2}$ suppresses considerably the effects of $\delta$ on the dynamics of the Universe evolution, and these terms wash-out for large $t$.
The trace anomaly (\ref{TraceFRW}) reads
 \[
 \langle T^{(QM)} \rangle=-\frac{3k_3}{4t^4}\,.
 \]
From $R=-\displaystyle{\frac{8\pi}{M_P^2}}\langle T^{(QM)} \rangle$ we find that
the parameter characterizing the heavy neutrino asymmetry (\ref{eta}) assumes the form
 \begin{equation}\label{eta1fin1}
  \eta = k_3 \frac{180}{\pi g_*} \sqrt{\frac{32^5 \pi^{15} g_*^5}{90^5}} \frac{T^{10}}{M M_P^9}\simeq
 2.5 k_3 \, 10^7 \frac{T^{9}}{M}\frac{1}{M_P^8} \,.
 \end{equation}
According to the leptogenesis scenario, the heavy neutrino asymmetry freezes at the decoupling temperature $T_D$ when the lepton-number violating interactions ($N_R \leftrightarrow N_R^c$) go out of equilibrium. The subsequent decays of these heavy neutrinos into the light standard model particles and the conversion of lepton asymmetry into baryon asymmetry can explain the observed baryon asymmetry of the Universe. In fact, for $T_D \sim (10^{16}-2\times 10^{16})$GeV and $M\sim (10^{9}-10^{12})$, respectively, one gets $\eta \sim 10^{-11}-10^{-10}$.

Let us finally compute the energy density of back-reaction of quantum fields and compare it with
the energy density of radiation. From Eqs. (\ref{Tmunu}) it follows
 \begin{eqnarray}\label{rhoQFT}
    \langle \rho\rangle &=& \langle T_{00}^{(QM)} \rangle = 18k_1\left(2{\ddot H} H +{\dot H}^2+10{\dot H} H^2\right) +3k_3 H^4 =\\
   & =  &  \frac{3k_3}{8t^4}\,. \nonumber
 \end{eqnarray}
The $k_1$-term vanishes identically. The total energy density is given by
 \begin{equation}\label{energytot}
    \rho= \rho_r + \langle \rho \rangle = \frac{\rho_0}{a^4}+\frac{{\cal A}_{qf}}{a^8}\,,\qquad {\cal A}_{qf}\equiv \frac{3k_3}{8}a_0^4\,.
 \end{equation}
where $\rho_r$ is the classical radiation defined in (\ref{rhoDM}).
The ratio between the energy densities $\langle \rho \rangle$ and $\rho_r$ reads
 \begin{equation}\label{ratio}
    r\equiv \frac{\langle \rho \rangle}{\rho_r}=\left(\frac{T}{T_*}\right)^4\,,
 \end{equation}
where we have definite $T_*$ as
 \[
 T_*\equiv \left[\frac{80}{k_3 \pi^4 g_* }\left(\frac{15}{16}\right)^2 \right]^{1/4}M_P \simeq \frac{1}{k_3^{1/4}}10^{18}\,\text{GeV}\,.
 \]
For temperatures $T< T_*$ we have that $r<1$, i.e. the energy density of quantum fields is subdominant with respect to the energy density of the radiation.
In particular, since the decoupling temperature of heavy neutrinos occurs at GUT scales, $T_D \sim 10^{16}$GeV, we infer $r\sim 10^{-9}\ll 1$ and the back-reaction is indeed subdominant over the radiation density.

\subsection{Gravitational Leptogenesis in Warm Inflation}

A further application of the gravitational leptogenesis scenario is the warm inflation \cite{Berera}
models as there is a large non-zero Ricci curvature from the inflaton potential during inflation and a
large temperature where the lepton number violating interaction can be at equilibrium.

Let us recall the central point underlying the warm inflation idea \cite{berera}. In the Inflationary dynamics the
scalar field carries most of the energy of the Universe. The inflaton however also interact with other fields,
but these interactions plays no role except to give rise to modifications to the effective scalar field through quantum
corrections. In the warm inflation scenario, instead, the effect of these interactions is not only to modify the scalar
field potential, but also to generate dissipation and fluctuation effects. In order that warm inflation works, it is required
that the time scale of quantum mechanical processes leading to the dissipation is much slower than the expansion rate
of the Universe (in such a way the whole system, inflaton and fields, would not equally distribute the available energy).

The Ricci scalar is related to the Hubble expansion rate during inflation as $R=-12 H^2$ and its time
derivative is related to the slow roll parameter $\epsilon =-\dot H/H^2$ as $\dot R=24 \epsilon
H^3 $. The lepton asymmetry (\ref{eta}) in warm inflation reads
 \begin{equation}
\eta_{wI} \simeq A_{wI} \frac{\epsilon H^3}{M_P\, T_{l} \, M}\,, \qquad A_{wI}\equiv \frac{180}{g_*}\simeq {\cal O}(1)\,.
\label{eta2warm}
 \end{equation}
$T_l$ is the light neutrino decoupling temperature (\ref{Tl}).

The power spectrum of curvature perturbation in thermal inflation
and the spectral index of scalar perturbations are expressed in
terms of $H$ and $\epsilon$ \cite{Moss}
 \begin{equation}\label{warminflpar}
 {\cal P}_{\cal R}=\left(\frac{\pi}{16}\right)^{1/2}\,
\frac{H^{1/2}\,\Gamma^{1/2}\,T}{M_P^2 \epsilon}\,, \quad
 n_s-1=-\frac{27}{4} \frac{H}{\Gamma} \epsilon\,,
 \end{equation}
where $\Gamma$ is the damping parameter in the inflaton equation
of motion and represents the coupling between the inflaton and the
thermal bath.
The WMAP observations \cite{wmap} provides the
amplitude of the curvature power spectrum and  of the spectral index, which are given by  ${\cal P}_{\cal R}=
(2.3 \pm 0.3) \times 10^{-9}$ and the spectral index $n_s=0.951 \pm 0.017$, respectively.

Combining (\ref{eta2warm}) and (\ref{warminflpar}) one can write $\epsilon$, $\Gamma$ and
$M$ in terms of $H$ and $T$:
 \begin{eqnarray}
    \Gamma &=& \left(\frac{27\sqrt{\pi}}{16}\right)^2 \frac{H^3T^2}{M_P^4}\frac{1}{{\cal P}_{\cal R}^2(1-n_s)^2}\,, \nonumber \\
    \epsilon &=& \frac{27\sqrt{\pi}}{16} \frac{H^2T^2}{M_P^4}\frac{1}{{\cal P}_{\cal R}^2(1-n_s)}\,, \label{HTwI} \\
    M &=& A_{wI}\frac{27\sqrt{\pi}}{16} \frac{H^5 T}{M_P^5 \eta_{wI}}\frac{1}{{\cal P}_{\cal R}^2(1-n_s)}\,,
 \end{eqnarray}
Choosing $H\simeq 8\times 10^{12}$GeV and $T \simeq 8 \times 10^{12}$GeV,  and using, consistently with WMAP data,
${\cal P}_{\cal R}\sim 2 \times 10^{-9}$ and $n_s=0.968$, the net baryon asymmetry $\eta_{wI}\simeq 10^{-10}$ in the warm inflation scenario follows
provided $\Gamma=7.1 \times 10^9$GeV, $\epsilon = 4.2\times 10^{-6}$, and $M=2.7 \times 10^{11}$GeV.
Finally, from (\ref{Tl}) one gets that $T_l\sim 10^{13}$GeV corresponds to the neutrino mass of $m_{\nu_3}=0.15$eV. Other models on
baryo/leptogenesis in warm Inflation scenario have been proposed in \cite{branderberger,bererawIbaryo}.


\section{Leptogenesis induced by spin-gravity coupling of neutrinos with the Primordial Gravitational Waves}
\setcounter{equation}{0}

The behavior of (relativistic) quantum systems in gravitational fields, as well as in inertial fields,
play a crucial role for investigating the structure of spacetime at the quantum level \cite{Audretsch}.
Quantum objects are in fact finer and more appropriate probes of structures that appear classically as results of
limiting procedures. On fundamental ground, it is expected that only a quantum theory of gravity will be able to provide a definitive answer
to questions regarding the fundamental structure of spacetime. However, the extrapolation of General Relativity from
ordinary terrestrial scales to Planck's scales is not free from subtle questions and new data coming from modern cosmology,
as discussed in the Introduction and previous Sections. Therefore,  one considers the gravitational background described
by General Relativity, hence considered as a classical filed, and matter as quantized fields propagating in a classical background \cite{birrell}.
In this respect, one is there considering considering the interaction of classical inertial and gravitational fields with
(relativistic) quantum objects.

Observations performed in \cite{COW} do confirm that both gravity and inertia interact with quantum systems
in ways that are compatible with General Relativity. This, per se, does not represent a test of General Relativity,
but shows that the effects of inertia and gravity on wave functions
are consistent with covariant generalization of wave equations dictated by General Relativity paradigms.
Clearly, to test fundamental theories a central role is played, at level of terrestrial experiments (Earth-bound and
near-space experiment), by inertial effects. Their identifications is therefore required with great accuracy and
represents a big challenge for future experiments. Inertial effects, on the other hands, provide a guide in the study of relativity
because, in all physical situations in which non-locality is not an issue,
the equivalence principle ensures the existence of a gravitational effect for {\it each} inertial effect.

Certainly the study of spin-gravity and spin-inertia coupling (as well as spin precession) represents a very active and relevant topics of
physics. Experiments of high energy physics indeed typically involve spin-1/2 particles and take place or in a gravitational environment or in non-inertial frames. Thanks to the progress of technology, for example, atomic interferometry and the physics of polarized
systems, the effects of the interactions of relativistic quantum particle with gravitation field, i.e. the spin-gravity coupling effects,
could be provide new insights of QFT in curved spacetime.
Spin-inertia and spin-gravity interactions and their effects in different physical situations are the subject of numerous theoretical (see for example
\cite{hehl115,hehl116,Audretsch,Huang,Mashhoon119,Mashhoon120,ryder,mashhoonGRG,Obukhov,Obhukovrev,mohantySat,adler1,Giudice,
papiniPRL,anandan,parker,Papinirev,oliveira,peres,silenko,Werner1990,bell,lambiaseIJMPD09,lambiaseApJ08,lambiaseMNRAS05,Lambiase:2005gt} and references therein) and
experimental efforts\cite{123,124,125,126,127}.
Spin precession in inertial and gravitational fields have been studies
in\cite{mas-ob,Lambiase:2012yi,Lambiase:2013ena,Lambiase:2013su,Papini:2001un,Lambiase:2004sm}.

In this Section we shall discuss the effect of spin gravity coupling in a cosmological context.
In particular, we are going to discuss the mechanism of Leptgenesis induced by spin-gravity coupling of neutrinos with the cosmological background \cite{Mohanty:2005ud,lambiaseNovaScience}.
The approach is based on QFT in curved spacetime, and in particular we write down the Dirac
equation in an expanding Universe. To this aim, we use the
vierbein formalism. We assume that the early phase of the Universe
is described by Inflation which generates the gravitational waves (tensor modes).
The latter split in the energy levels of (Majoarana) neutrinos and antineutrinos, which ultimately results in the
creation of a lepton asymmetry in the presence of lepton number violating interactions. This mechanism gives rise to
the generation of a net leptogenesis.

\subsection{Dirac equation in curved space-time and the fermion dispersion relation}\label{SectionSpin}

In passing from flat to curved space time we use the standard prescription
$\partial \to \nabla$, and $\eta_{\mu\nu} \to g_{\mu\nu}$.
The procedure to replace flat space tensors with "curved space"
tensor cannot be extended to the case of spinors. This procedure
works with tensors because the tensor representations of $GL(4,
R)$, i.e. the group of $4\times 4$ real matrices, behave like
tensors under the subgroup $SO(3,1)$. Thus, considering the vector
representation, as an example, one gets
 \[
 V^{\prime \, \mu}(x')=\frac{\partial x^{\prime\, \mu}}{\partial x^\nu}
 V^{\nu}(x) \quad
 \stackrel{ x'=\Lambda^\mu_{\,\,
\nu}x^\nu }{\longleftrightarrow} \quad  V^{\prime \,
\mu}(x')= \Lambda^\mu_{\,\, \nu} V^{\nu}(x)
 \]
But there are no representation of $GL(4,R)$ which behave like
spinors under $SO(3,1)$, i.e. there not  does exists a function of $x$ and
$x'$ which reduces to the usual spinor representation of the
Lorentz group ($D(\Lambda)$) for $x'=\Lambda^\mu_{\,\, \nu}x^\nu$.
Therefore, to write down the general covariant coupling of spin-1/2 particles
to gravity, we have to use the vierbein formalism.

A vierbein fields (or tretad) is defined as
 \[
 e^a_{\,\, \mu}(X)=\frac{\partial \xi^a(x)}{\partial x^\mu}\vert_{x=X}
 \]
where $\xi^a$  are the local inertial coordinate,  $x^\mu$ the generic coordinate,
$\xi^a(x)\to \xi^{\prime \, a}(x)=\Lambda^a_b(x) \xi^b(x)$, and  $\Lambda^T\eta\Lambda = \eta$.

The quantities $e^a_{\,\, \mu}(x)$ constitute a set of four coordinate vectors
which form a basis for the (flat) tangent space to the curved
space at the point $x=X$. Under the coordinate transformation $x\to x'=x(x)$, the vierbeins
$e^a_{\,\, \mu}(x)$ transform as (see Table \ref{ta3})
 \[
 e^{\prime\, a}_{\phantom{\prime a} \mu}(x^\prime)=
\frac{\partial x^\nu}{\partial x^{\prime \, \mu}}\, e^a_{\,\,
\nu}(x)\,, \quad (\xi^{\prime\, a}(x^\prime)=\xi^{a}(x))\,.
 \]

\begin{table}[ph]
\caption{Vierbeins transformations}
{\begin{tabular}{@{}cccc@{}} \toprule
 Under coordinate transformations     &  Under  local Lorentz transformations    \\
  the vierbeins $e^a_{\,\, \mu}(x)$ transform as   & the vierbeins   $e^a_{\,\, \mu}(x)$ transform as  \\
  & \\
  $ e^{\prime\, a}_{\phantom{\prime a} \mu}(x^\prime)=
 \frac{\partial x^\nu}{\partial x^{\prime \, \mu}}\, e^a_{\,\,
 \nu}(x) $  &  $e^{\prime\, a}_{\,\,\,\, \mu}(x^\prime)= \Lambda^a_{\,\, b}(x)\, e^b_{\,\,
 \nu}(x)$   \\
   & \\
  $ \xi^{\prime\, a}(x^\prime)=\xi^{a}(x)$  & whereas $x^\mu$ does not transform \\  \botrule
\end{tabular} \label{ta3}}
\end{table}

The metric $g_{\mu\nu}(x)$ is related to the vierbein fields by
the relation $g_{\mu\nu}(x)=\eta_{ab}\,\,e^a_{\,\, \mu}(x)e^b_{\,\, \nu}(x)$, where
$\eta_{ab}$ is the Minkowsky metric (in the local inertial frame). It follows
 \[
 \delta^\mu_\nu=e_a^{\,\, \mu}(x)e^a_{\,\, \nu}(x)\quad
 \mbox{i.e. $e^a_{\,\, \nu}(x)$ is the inverse of $e_a^{\,\,
\nu}(x)$} \quad \to \quad \eta^{ab}=g^{\mu\nu}(x)e^a_{\,\,
\mu}(x) e^b_{\,\, \nu}(x)\,.
 \]
Spinor fields are coordinate scalars which transforms under local
Lorentz transformations as $\psi_\alpha(x) \to \psi^\prime_\alpha(x)=D_{\alpha\beta}[\Lambda(x)] \psi_\beta(x)$,
where $D_{\alpha\beta}[\Lambda(x)]$ is the spinor representation
of the Lorentz group and $\psi_\alpha$ is the component of the
spinor $\psi$ (not be confused with general coordinate indices).
Since $\Lambda$ {\it does depend} on $x$, $\partial_\mu
\psi_\alpha$ {\it does not} transform like $\psi_\alpha$ under
local Lorentz transformations. To obtain a Lagrangian invariant
under generic coordinate transformation one has to define the
covariant derivative $
 D_\mu\psi_\alpha\equiv \partial_\mu
 \psi_\alpha-[\Omega_\mu]_{\alpha\beta} \psi_\beta$,
where $[\Omega_\mu]_{\alpha\beta}$ is the connection matrix.
Therefore one requires
$D_\mu\psi_\alpha\to D_{\alpha\beta}[\Lambda(x)]\, D_{\mu} \psi_\beta(x)$,
provided
 \[
 \Omega^\prime_\mu=D(\Lambda)\Omega_\mu D^{-1}(\Lambda)-(\partial_\mu
 D(\Lambda))^{-1} D^{-1}(\Lambda)
 \]
The connection matrix $[\Omega_\mu]_{\alpha\beta}(x)$ can be
written as
 \[
[\Omega_\mu]_{\alpha\beta}(x)=\frac{i}{2}[S_{ab}]_{\alpha\beta}\omega_\mu^{\phantom{\mu}ab}(x)
 \]
where $S_{ab}=\frac{\sigma_{ab}}{2}=\frac{i[\gamma_a,
\gamma_b]}{2}$ are the generators of the the Lorentz group in the
spinor representation and $\omega^{\,\,\,a}_{\mu\,\, b}$ the spin
connections. The spinor representation of the Lorentz group can be
written as $D[\Lambda(x)]=\exp[-(i/2)S_{ab}\theta^{ab}]$. The
covariant derivative acts on vierbeins as
 \[
D_\mu e^a_{\,\, \nu}=\partial_\mu e^a_{\,\,
 \nu}-\Gamma^\lambda_{\mu\nu}e^a_{\,\,
 \lambda}-\omega^{\,\,\,a}_{\mu\,\, b}e^b_{\,\, \nu}
 \]
and the condition $D_\mu e^a_{\,\, \nu}=0$ allows to determine the
spin connections
 \[
 \omega_{bca} =
 e_{b\lambda}\left(\partial_a e^\lambda_{\;\;c} + \Gamma^\lambda_{\gamma
 \mu} e^\gamma_{c} e^\mu_a \right)\,.
 \]
Therefore, the  general covariant coupling of spin $1/2$ particles
to gravity is given by the Lagrangian
 \begin{equation}\label{L-Dirac}
{\cal{L}} = \sqrt{-g} (\bar{\psi} \gamma^a D_a \psi - m \bar{\psi}
\psi )
 \end{equation}
where $D_a =  \partial_a - \frac{i}{4} \omega_{bca} \sigma^{bc}$
is the covariant derivative before introduced. The Lagrangian
is invariant under the local Lorentz transformation of the
vierbein and the spinor fields.
By using the Dirac matrices properties
\[
 \gamma^a [\gamma^b,
 \gamma^c]=\eta^{ab}\gamma^c+\eta^{ac}\gamma^c-i\varepsilon^{dabc}\gamma_g\gamma^5
\]
the Lagrangian density (\ref{L-Dirac}) can be written in the form
 \begin{equation}\label{L-B}
 {\cal L}= \det(e)~\bar \psi \left(~ i \gamma^a \partial_a~
 -~m ~ - ~ \gamma_5 \gamma_d
 B^d~  \right) \psi\,,
 \end{equation}
where
 \begin{equation}\label{Bexpression}
 B^d = \epsilon^{abcd} e_{b \lambda}  (\partial_a
e^{\lambda}_c + \Gamma^\lambda_{\alpha \mu} e^\alpha_c e^\mu_a)\,.
 \end{equation}
In a local inertial frame of the fermion, the effect of a
gravitational field appears as a {\it axial-vector interaction}
term shown in ${\cal L}$.

\subsection{Neutrinos effective Lagrangian in a local inertial frame}

To determine the dispersion relation of neutrinos propagating in a {\it perturbed} FRW Universe, we have to
compute $B^d$. Perturbations are generated by
quantum fluctuations of the inflaton. Notice that for a FRW
Universe the $B^a$-term vanishes due to symmetry of the metric.
The general form of perturbations on a flat FRW
expanding universe can be written as
 \[
 ds^2=a(\tau)^2 [(1+2 \phi) d\tau^2 - \omega_i dx^i d\tau -
 ((1+ 2\psi) \delta_{ij} + h_{ij}) dx^i dx^j]
 \]
where $\phi, \psi$ are scalar fluctuations, $\omega_i$ the vector
fluctuations and $h_{ij}$ the tensor fluctuations of the metric.
Of the ten degrees of freedom in the metric perturbations only six
are independent and the remaining four can be set to zero by
suitable gauge choice. We work in the TT gauge: $h^i _i=0$,
$\partial^i h_{ij}=0$. In the TT gauge the perturbed FRW metric
can be expressed as
 \[
 ds^2=a(\tau)^2 [(1+2 \phi) d\tau^2 - \omega_i dx^i d\tau  -
 (1+ 2\psi  - h_{+}) dx_1^2
 -(1+ 2\psi+h_{+})dx_2^2
 \]
 \[
 - 2 h_\times dx_1 dx_2 -(1+ 2\psi) dx_3^2].
 \]
An orthogonal set of vierbiens $e^a_\mu$ for this metric is given
by
 \begin{eqnarray}\nonumber
 e_{\mu}^a  = a(\tau) \begin{pmatrix}
  1+ \phi& -\omega_1 & -\omega_2 & -\omega_3 \cr
  0 & -(1+ \psi)+h_+/2  &  h_\times & 0\cr
  0 & 0 & -(1+  \psi)-h_+/2 & 0 \cr
  0& 0& 0& -(1+ \psi)\,\end{pmatrix}\,.
 \end{eqnarray}
 For our application we need only the tensor perturbations. Explicit calculations give
$B^a=(\partial_\tau h_\times, 0, 0, \partial_\tau h_\times)$.
Using $\psi=(\nu_L, \nu_R)^T$ into (\ref{L-Dirac}), one gets
 \[
  {\cal L}=
  \det(e) [\left(i{\bar \nu_L}\gamma^a \partial_a\nu_L+i{\bar
  \nu_R}\gamma_a \partial_a \nu_R \right) +
    m {\bar \nu_L} \nu_R  +m^\dagger  {\bar \nu_R} \nu_L +
    \]
    \[
   + B^a (\bar \nu_R \gamma_a \nu_{R} - \bar
   \nu_{L}\gamma_a\nu_L)]\,.
 \]
The fermion bilinear term $\bar \psi
\gamma_5 \gamma_a \psi $ is odd under $CPT$ transformation. When
one treats $B^a$ as a background field  then the interaction term
in ${\cal L}$ explicitly violates $CPT$. When the primordial
metric fluctuations become classical, i.e  there is no
back-reaction of the micro-physics involving the fermions on the
metric and $B^a$ is considered as a fixed external field, then
$CPT$ is violated spontaneously. Moreover, we consider only
Standard Model fermions, so that $L_\nu=+1$ for neutrinos and
$L_{\bar \nu}=-1$ for antineutrinos ($L$ is lepton number), and we
consider Majorana spinors $\nu_R = (\nu_L)^c$, i.e. $\nu_R$ is the
charge-conjugate of $\nu_L$. With this choice, the mass term in
${\cal L}$ is of the Majorana type (the generation index is
suppressed).

\subsection{The Leptogenesis mechanism from spin-gravity coupling following Inflation}

According to the general setting, one has to assume that there are GUT processes that violate lepton number above
some decoupling temperature $T_D$. Results of Section \ref{Section4} imply that
in the presence of non-zero metric fluctuations, there is a split in energy levels of $\nu_{L,R}$ (a different effective
chemical potential), so that the
dispersion relation of $\nu_{L,R}$ fields reads (see also\cite{mukhop})
$E_{L,R}(p) = E_p  \mp \displaystyle{\left(B_0 -\frac{{\bf p} \cdot {\bf B}}{p}\right)}$, where as usual $E_p =p+m^2/2p$ and $p=|{\bf p}|$.

The equilibrium value of lepton asymmetry generated for all $T>T_D$ turns out to be
(Eq. (\ref{nchiralitypart}))
 \begin{equation}\label{DeltanGW}
 \Delta n = \frac{g T^3}{6} \left(\frac{B^0}{T}\right)\,.
 \end{equation}
where $\Delta n \equiv n(\nu_L)-n(\nu_R)$ and $p \gg m_\nu $ and $B_0 \ll T$ has been used.
The dependence on ${\bf B}$ drops out after angular integration

To evaluate $B^0$, we need to compute the spectrum of
gravitational waves $h({\bf x},\tau)$ during inflation. To this aim, we express $h_{\times}$
in terms of the creation- annihilation operator
 \begin{equation}\label{h}
 h({\bf x}, \tau)=\frac{\sqrt{16 \pi}}{a M_p}\int \frac{d^3 k}
 {(2\pi)^{3/2}}
 \left(a_{\bf k} ~f_k(\tau) +a^{\dagger}_{-{\bf k}}~f_{k}^*(\tau)
 \right)e^{i{\bf  k} \cdot {\bf x}}\,,
 \end{equation}
where ${\bf k}$ is the comoving wavenumber, $k=|{\bf k}|$, and
$m_P= 1.22 \,\,10^{19}GeV$ is the Planck mass. Remembering that
$B^0=\partial_\tau h$ one gets that the two point correlation
function for $B^0$ is (see \ref{PWGW} for details)
 \[
 \langle B^0({\bf x}, \tau), B^0({\bf x}, \tau) \rangle =
 \int \frac{dk}{k}\left(\frac{k}{a}\right)^2(h_k^{rad})^2
 =\frac{4}{\pi}\left(\frac{H_I}{m_P^2}\, T^2 1.67
 \sqrt{g_*}\right)^2 N
 \]
Therefore, the r.m.s value of spin connection that determines the
lepton asymmetry $\Delta n$ is
 \[
 (B_0)_{rms} \equiv  \sqrt{ \langle B_0^2\rangle} = \frac{2
}{\sqrt{\pi}}\left(\frac{H_I}{m_P^2} \, \, T^2\, 1.67\,
\sqrt{g_*}\right)\,\sqrt{N}\,.
 \]
The lepton asymmetry $\Delta n$  as a function of temperature can
therefore be expressed as (considering three neutrino flavors)
 \[
  \Delta n(T) = \frac{gT^3}{6}\frac{(B_0)_{rms}}{T}
=\frac{1}{\sqrt{\pi}}\,(1.67\, \sqrt{g_*})\, \sqrt{N}
 \left(\frac{T^4\, H_I} {m_P^2}\right)\,.
 \]
The lepton number to entropy density ($s=0.44~ g_*~ T^3$)
is therefore
 \[
 \eta \equiv \frac{\Delta n(T)}{s(T)}
 \simeq 2.14 \,  \frac{T\,H_I \,\sqrt{N}}{ m_P^2\,\sqrt{g_*}}
 \]
The lepton asymmetry is evaluated at the decoupling temperature
$T_D$. According to general setting, lepton number asymmetry will
be generated as long as the lepton number violating interactions
are in thermal equilibrium. Once these reactions decouple at some
decoupling temperature $T_D$, which we shall determine, the
$\Delta n(T)/s(T)$ ratio remains fixed for all $T<T_D$. To
calculate the decoupling temperature of the lepton number
violating processes we turn to a specific effective dimension five
operator which gives rise to Majorana masses for the neutrinos (\ref{dim5})
The $\Delta L=2$ interactions that result
from the dimension five operator (\ref{dim5}) are
given in (\ref{ints}).
In absence of GWs, it follows that the forward reactions are equal
to the backward reactions, and therefore no net lepton number
is generated. On the contrary, in presence of GWs, the
forward reactions are different by backward reactions, and the
energy levels of the left and right helicity neutrinos are no
longer degenerate, $E_L \neq E_R$, which implies a difference in
the number density of left and right handed neutrinos (at thermal
equilibrium) $\Delta n = n(\nu_L)-n(\nu_R) \neq 0$.
This process continues till the $\Delta L=2$ interactions
decouple.

Next step now is to calculate the decoupling temperature $T_D$ by using (\ref{H=T}).
The interaction rate for the interaction $\nu_{L
\alpha} +\phi^0 \leftrightarrow \nu_{R \beta} +\phi^0$ is (see (\ref{Gamma1}))
 \[
  \Gamma =\langle n_\phi~ \sigma\rangle =
\frac{0.122}{\pi} \frac{|C_{\alpha\beta}|^2 T^3}{M^2}
 \]
In the electroweak era, when the Higgs field in ${\cal{L}}_W$
acquires a $vev$, $\langle\phi\rangle =(0,v)^T$ (where
$v=174~GeV$), the five dimensional Weinberg operator gives rise to
a neutrino mass matrix $m_{\alpha \beta}= \frac{v^2 C_{\alpha \beta}}{M}$.
This implies that in our calculations, we can substitute the
couplings $\frac{C_{\alpha \beta}}{M}$ in terms of the light left
handed Majorana neutrino mass, i.e. $\frac{C_{\alpha \beta}}{M} \to \frac{m_{\alpha\beta}}{v^2}$.
The decoupling temperature $T_D$ turns out to be
 \[
 T_D= 13.68 \pi~ \sqrt{g_*}~ \frac{v^4}{m_\nu^2 \, m_{P}}\,,
 \]
where $m_\nu$ is the mass of the (heaviest) neutrino. Substituting
$T_D$ into the expression for $\eta$, we finally obtain the
formula for lepton number
 \begin{equation}
 \eta = 92.0 \,\left(\frac{v^4 \, H_I}{m_\nu^2 \, m_P^3}\,\right)\,\sqrt{N}\,. \label{etafinalae2013}
 \end{equation}
The input parameters we used for our estimations are: $a)$ The amplitude of the $h_\times\sim 10^{-6}$ or equivalently the
curvature during inflation $H_I \sim 10^{14}$GeV or the scale of
inflation is the GUT scale, $V^{1/4} \sim 10^{16}$GeV which is
allowed by CMB \cite{pilo}.
$b)$ Neutrino Majorana mass in the atmospheric neutrino scale\cite{Vogel,super-K}
 $m_{\nu}^2 \sim 10^{-3}$eV$^2$.
$c)$  Duration of inflation $H_I t =N \sim 100$ (needed to solve the horizon and entropy problems in the
standard inflation paradigm, Eq. (\ref{durationofInflation})).
The parameters entering (\ref{etafinalae2013}) are well within experimentally acceptable
limits. The magnitude of baryogenesis is therefore
 \begin{equation}\label{finaleetaGW}
\eta = 7.4 \, 10^{-11} \frac{H_I}{4 \, 10^{14} GeV}\,\frac{2.5 \,
10^{-3} eV^2}{m_\nu^2}\,\frac{\sqrt{N}}{10}
   \sim 10^{-11} - 10^{-10}\,,
 \end{equation}
which is compatible with the previous values (\ref{etaCMB}) and
(\ref{etaBBN}). According to \cite{fukugita}, a lepton asymmetry generated at an earlier epoch
gets converted to baryon asymmetry of the same magnitude by the
electroweak sphalerons.

The mechanism here discussed makes use of the standard QFT in
curved space-time, which gives rise to the conventional spin gravity
coupling of neutrinos with the gravitational background. This
leads to the coupling of axial-vector current with the four-vector
$B^a$ which accounts for the curved background. As we have seen,
left-handed and right-handed fields couple differently to gravity
and therefore have different dispersion relations
(the equivalence principle is hence violated). Moreover, the model
uses the Majorana neutrinos because one needs of a violation of
lepton number which generates the lepton asymmetry. In order that spin-gravity coupling of neutrinos with
the gravitational background be a viable model, one has to assume that
the early Universe is described by Inflation. Remarkably, no free parameters are present in the final expression
for the lepton asymmetry given by (\ref{etafinalae2013}).

\subsection{CPT Violation in an Expanding Universe -  String Theory}

Recently Ellis, Mavromatos and Sarkar\cite{ellis,mavromatos} proposed a model in which they explore the possibility
to violate CPT in an expanding Universe
in the framework of String Theory, and generate a net baryon/lepton asymmetry via Majorana neutrinos.

The model goes along the following line. According to String Theory, besides to the spin-2 graviton
(described by the usual symmetric tensor $g_{\mu\nu}$), the theory contains also the spin-0 dilaton field (described by the scalar field $\Phi$)
and the anti-symmetric tensor Kalb-Ramond field (described by $B_{\mu\nu}$). The latter enters into the effective action via the field
$H_{\mu\nu\rho}=\partial_{\mu}B_{\nu\rho}+p.c.$, where $p.c.$ stands for the cyclic permutation of the indices $\{\mu,\nu,\rho\}$
($H_{\mu\nu\rho}$ plays the role of torsion\cite{gross}). The effective action reads (in the Einstein frame)
 \begin{equation}\label{stringaction}
    S = \frac{M_s^2 V_c}{16\pi}\int d^4 x \sqrt{-g} \left(R(g)-2\partial_\mu \Phi \partial^\mu \Phi -\frac{e^{-4\Phi}}{12} H_{\mu\nu\rho}H^{\mu\nu\rho}
    + {\cal O}(\alpha^\prime)\right)\,.
 \end{equation}
Here $M_s=1/\sqrt{\alpha^\prime}$ represents the string scale mass, $V^c$ the compactification volume which is expressed, together with the compact
radii, in terms of the $\sqrt{\alpha^\prime}$ units. In this model ne gets that the connections are generalized as
 \begin{equation}\label{connectiongener}
    {\bar \Gamma}^\lambda_{\mu\nu}=\Gamma^\lambda_{\mu\nu}+ T^\lambda_{\mu\nu}\,,
 \end{equation}
where $\Gamma^\lambda_{\mu\nu}$ are the usual Christoffel symbols, and
 \begin{equation}\label{torsionST}
    T^\lambda_{\mu\nu} \equiv e^{-2\Phi}H^\lambda_{\mu\nu} = -T^\lambda_{\nu\mu}\,.
 \end{equation}
The four-vector $B^d$ given in (\ref{Bexpression}) now reads
  \begin{equation}\label{BexpressionST}
 B^d = \epsilon^{abcd} e_{b \lambda}  (\partial_a
e^{\lambda}_c + {\bar \Gamma}^\lambda_{\alpha \mu} e^\alpha_c e^\mu_a)\,.
 \end{equation}
The anti-symmetric tensor can be written in terms of the pseudo-scalar axion-like field $b(x)$
\begin{equation}\label{h=b}
H_{\mu\nu\rho}=e^{2\Phi}\epsilon_{\mu\nu\rho\sigma}\partial^\sigma b\,,
\end{equation}
where $\epsilon_{0123}=\sqrt{-g}$. Field equations of String theory provides the solution \cite{antoniadis}
\begin{equation}\label{b-solution}
    b(x)=\sqrt{2}e^{-\Phi_0} \sqrt{Q^2} \frac{M_s}{\sqrt{n}}\, t\,,
\end{equation}
where $\Phi_0$ is a constant appearing in the time evolution of the dilaton field $\Phi(t)=-\ln t+\Phi_0$,
$Q^2>0$ is the central charge deficit and $n$ is an integer associated to the Kac-Moody algebra of the
underlying world sheet conformal field theory. According to (\ref{connectiongener})-(\ref{b-solution}) one finds
that the non vanishing component of $B^d$ is
 \begin{equation}\label{B0-ST}
    B^0=\epsilon^{ijk}T_{ijk}=6\sqrt{2Q^2}\, e^{-\Phi_0} \frac{M_s}{\sqrt{n}}\, \text{GeV}\,.
 \end{equation}
The net baryon asymmetry between the (Majorana) neutrino-antineutrino is
\begin{equation}\label{asymmetryST}
    \eta \sim \frac{B_0}{T_d}\sim \frac{e^{-\Phi_0}M_s\sqrt{Q^2}}{\sqrt{n}T_d}\,.
\end{equation}


\subsection{Leptogenesis induced by Einstein-Cartan-Sciama-Kibble torsion field}

Another interesting model of matter-antimatter asymmetry has been proposed by Poplawski \cite{poplawski}.
It is based on Einstein-Cartan-Sciama-Kibble (ECSK) theory of gravity \cite{cartan} in which the usual Hilbert-Einstein action
(\ref{hilbert}) incorporates the  torsion field. The latter therefore extend General Relativity to include matter
with intrinsic spin-1/2, which produce torsion, and provides a more general theory of local gauge with respect to the
Poincare group \cite{poplawski1} (interesting applications of ECSK can be find in \cite{allECSK,Capozziello:2001mq} and references therein).

Spinors coupled to the torsion fields evolves according to Helh-Datta equation \cite{datta}
 \begin{equation}\label{dattaeq}
    i e_a^{\phantom{a}\mu} \gamma^a \nabla_a\psi - m \psi = -\frac{3}{8M_P^2}({\bar \psi}\gamma^5 \gamma_a \psi)\gamma^5 \gamma^a \psi\,,
 \end{equation}
where $\nabla_a$ represents the covariant derivative with respect to the affine connection (Christoffel connections). The corresponding equation
for the charge conjugate (C) spinor $\psi^{\text{C}}$ is
 \begin{equation}\label{dattaeqCC}
    i e_a^{\phantom{a} \mu} \gamma^a \nabla_a \psi^{\text{C}} - m\psi^{\text{C}} = +\frac{3}{8M_P^2}({\bar \psi^{\text{C}}}\gamma^5 \gamma_a \psi^c)\gamma^5 \gamma^a \psi^{\text{C}}\,,
 \end{equation}
The equations (\ref{dattaeq}) and (\ref{dattaeqCC}) are therefore different, leading to a different shift of energy spectrum, generating in such a
way an asymmetry. In fact, the energy levels for a free fermion ($X$) and antifermion (${\bar X}$) resulting from ECSK theory are (in the ultrarelativistic
limit)
  \begin{equation}\label{ensplitECSK}
    E_X=p+\alpha \kappa N\,, \qquad E_{\bar X}=p-\alpha \kappa N\,,
  \end{equation}
where $\alpha$ is a numerical factor of the order of unity, and $N$ is the inverse normalization of Dirac spinor (it is of the order of $N\sim E^3\sim T^3$).
Here $X$ and ${\bar X}$ refer to heavy fermion carrying baryon and antibaryon number, respectively, and are dubbed archeons and antiarcheons. They
candidate for a possible component of DM.
Eqs. (\ref{sentropy}) and (\ref{nchiralitypart}) then imply that the net baryon asymmetry is
$\eta \sim T_D^2/M_P^2$, where $T_D$ is the decoupling temperature that must assume the value $T_D \sim 10^{13}$GEV in order to generate the observed
baryon asymmetry. One can estimate the mass $m_X$ of the (anti)archeon by equating the decay rate $\Gamma\sim \displaystyle{\frac{G_F^2 m_X^5}{192\pi^3}}$
with the expansion rate of the Universe during the radiation era, Eq. (\ref{exprateH}): $m_X =m_{\bar X}\sim 10$TeV. This value is of the same order of
magnitude of the maximum energy of a proton beam at LHC ($\sim 7$TeV).

\section{Conclusion}

Understanding how the baryon asymmetry of the universe originated is one of the fundamental goals of modern cosmology.
As we have seen, particle physics, as well as cosmology, have provided with a number of possibilities.
They involve very fascinating physics, but with varying degrees of testability. In fact, all the baryogenesis models are indeed able
to derive the correct estimation of $\eta\sim 10^{-10}$, but it is rather difficult, if not impossible, to exclude or confirm one or other
scenario.

A possibility to discriminate among the plethora of baryogenesis models is to investigate its predictibility or
compatibility with certain form of DM (besides the models discussed in this review, see also\cite{baryo+DM}).
In other words, one should expect that a realistic model of baryogenesis is able to determine both the right value of $\eta$,
and to explain the magnitude of the ratio $\rho_B/\rho_{DM}$ fixed by cosmological observations.
An example is provided by the Affleck-Dine Baryogenesis: The model explains the observed baryon asymmetry of the Universe,
and supersymmetric particles are favored candidates of DM in the Universe. In this respect, as discussed in this review,
gravitational baryogenesis  represents also an interesting framework. In fact  because if from one side, in these models
one is able to recover the correct estimation of the parameter $1eta$, from the other side they are compatible
with the present cosmological data of an epanding Universe, and hence the necessity to invoke new form of energy or matter,
hence DM and DE. This is the case of $f(R)$ gravity, but also variant of these models and more generally, scalar tensor theories
are available candidates.

A question that arises is whether the universe is baryon-symmetric on cosmological
scales, and eventually separated into domains which are either dominated by baryons
or antibaryons. One then would expect to detect, due to annihilations,
an excess of gamma rays. There are no evidences of the existence of such a cosmic anti-matter.
In fact, the analysis of $p{\bar p}$ annihilation in gamma rays ($p{\bar p} \to \pi^0\to 2\gamma$), with $E_\gamma\sim 100$MeV,
allows to conclude that the nearest rich antimatter region (anti-galaxy) should be away from at distance\cite{rujula,kinney}
$D\geq (10-15)$Mpc. This results indicate hence that the patches of matter-antimatter should be
as large as the presently observable Universe. However, no mechanism is known which is able to explain how to separate out these domains of matter-antimmater.
There is anyway an existing as well as many planned  experimental activity in searching for
cosmic antimatter\cite{PAMELA,ATIC7,Fermi-LAT8,Hess}.

It is still unclear to cosmologists and particle physicists what scenario was realized in nature to generate the observed
baryon asymmetry in the Universe.  What has arisen in the last years is that no model of Baryogenesis is complete without incorporating
the idea s underlying the Leptogenesis.  This is also supported by recent results of neutrino physics.
LHC experiment and the new generation of linear colliders will certainly allow a deeper understanding and a considerable progress of this fundamental
problem.  In any case, the baryon asymmetry in the Universe furnishes a clear evidence that a new physics,
beyond the Standard Model of particle physics and the Standard Cosmological model, is called for.

\section*{Acknowledgments} The authors thank Prof. Dh. Ahluwalia and Prof. E.H. Chion for invitation to write this review.
G. Lambiase thanks the University of Salerno and ASI (Agenzia Spaziale Italiana) for partial support through the
contract ASI number I/034/12/0.

\appendix

\section{Out equilibrium condition}\label{outequ}

For completeness, we discuss on a different setting, the departure from thermal equilibrium.

If all particles in the Universe were in thermal equilibrium then there would be no preferred direction for the time $T$
and also if $B$ asymmetry could be generated, it would be prevented by $CPT$ invariance.
Therefore, also the violation of the $CP$ symmetry would be irrelevant.

Consider a species of massive particle $X$ in thermal equilibrium at temperatures $T\ll m_X$. Let $m_X$ its mass.
The number density of these particles is
 \[
 n_X\approx g_X (m_X T)^{3/2}e^{-\frac{m_X}{T}+\frac{\mu_X}{T}}\,,
 \]
where, as usual, $\mu_X$ indicate the chemical potential. The species $X$ are in chemical equilibrium
if the rate $\Gamma_{inel}$ of inelastic scatterings (responsible of the variation of the number of $X$ particles in the plasma,
according to the processes $X+A \to B+C$) is larger than the expansion rate of the Universe
$\Gamma_{inel} \gg H$. This allows to write a relation among the
different chemical potentials of the particles involved in the process
$\mu_X+\mu_A=\mu_B+\mu_C$. The number density of the antiparticle ${\bar X}$ which have the same mass
of particles $X$, $m_{\bar X}=m_X$, but opposite chemical potential $\mu_{\bar X}=-\mu_X$ due to the process
$X+{\bar X}\to \gamma+\gamma$ with $\mu_\gamma=0$, is
\[
 n_{\bar X}\approx g_X (m_X T)^{3/2}e^{-\frac{m_X}{T}-\frac{\mu_X}{T}}
 \]
If the $X$ particle carries baryon number, then $B$ will get a contribution from
 \[
 B\propto n_X-n_{\bar X}=2g_X (m_XT)^{3/2}e^{-\frac{m_X}{T}}\sinh\frac{\mu_X}{T}\,.
 \]
If there exist $B$-violating reactions (first Sakharov's condition) for the species $X$ and ${\bar X}$, such as $X+X\to {\bar X}+{\bar X}$,
then the chemical potential is zero, $\mu_X=0$. As a consequence, also the relative contribution of the $X$ particles to the net baryon
number vanishes. Therefore, only a departure from thermal equilibrium can allow for a finite baryon excess, that means that
the form of $n_{X, {\bar X}}$ has to be modified,

The typical example of the out-of equilibrium decay can be represented by the following steps:
Let $X$ be a heavy particle such that $m_X > T$ at the decay time, and let $X\to Y+B$ the decay process.
When the energy of the final state is given by $E_{Y+B}\sim {\cal O}(T)$, then there is no phase space for the inverse decay to occur.
The final state $Y+B$ does not have enough energy to create a heavy particle $X$
(the rate for $Y+B\to X$ is Boltzmann suppressed, i.e. $\Gamma(Y+B\to X)\sim e^{-m_X/T}$).

\section{The physics of Sphalerons}\label{SphPhysics}

In the EW theory, the most general Lagrangian invariant under the
SM gauge group and containing only color singlet Higgs fields is
automatically invariant under global abelian symmetries. The latter are associated to
the baryonic and leptonic symmetries. It is hence not
possible to violate $B$ and $L$ at tree level, as well as in any order of perturbation theory.

The perturbative expansion, however, does not describe  all the dynamics of the theory. 't Hooft provided
 a scenario in which nonperturbative effects (instantons) may give rise
to processes which the combination $B+L$ is violated, whereas the (orthogonal) combination $B-L$ does not.
In some circumstances, such as  the early Universe at very high temperature,
the processes that violate the  baryon and lepton number may be fast enough. These processes may be
significant role for baryogenesis mechanisms.

At the quantum level, the baryon and lepton symmetries are anomalous (triangle anomaly)
 \[
 \partial J_B^\mu=\partial_\mu J^\mu_L=
 n_f\left(\frac{g^2}{32\pi^2}W^a_{\mu\nu}{\tilde W}^{a\mu\nu}-\frac{g'^2}{32\pi^2}
 F_{\mu\nu}{\tilde F}^{\mu\nu} \right)
 \]
where $g, g'$ are the gauge coupling of $SU(2)_L$ and $U(1)_Y$, $n_f$ the number of families,
 \[
{\tilde W}^{a\mu\nu}=\frac{1}{2}\epsilon^{\mu\nu\alpha\beta}W^a_{\alpha\beta}
\]
the dual of $SU(2)_L$ field strength tensor,
 \[
 {\tilde F}^{\mu\nu}=\frac{1}{2}\epsilon^{\mu\nu\alpha\beta}F_{\alpha\beta}
 \]
the dual of $U(1)_Y$ field strength tensor.
The change of baryon number, which is closely related to the
vacuum structure of the theory, is given by
 \[
\Delta B=B(t_f)-B(t_i)=\int_{t_i}^{t_f}\int d^3x \partial^\mu J_\mu^B= n_f[N_{CS}(t_f)-N_{CS}(t_0)] \,,
 \]
with
 \[
 N_{CS}(t)=\frac{g^2}{32\pi^2}\int d^3x \epsilon^{ijk}Tr\left(A_i\partial_j A_k+\frac{2}{3}i g A_i A_j A_k\right)\,.
 \]
$N_{CS}$ is the Chern-Simons number. For vacuum to vacuum transition, the field $A$ represent a (pure) gauge
configuration, whereas the Chern-Simons numbers $N_{CS}(t_f)$ and $N_{CS}(t_i)$ assume integer values.
In a non-abelian gauge theory, there are infinitely many degenerate ground states (labeled by the Chern-Simons number
$\Delta N_{CS}=\pm 1, \pm 2, \pm 3, \ldots$).
In field space, the corresponding point are separated by a potential barrier. The {\it height} of his barrier gives the sphaleron
energy $E_{sp}$. Because the anomaly, jumps in the Chern-Simons numbers are associated with changes of baryon and lepton number
$\Delta B= \Delta L=n_f \Delta N_{CS}$. The smallest jump in the Standard Model is characterized by
$\Delta B= \Delta L=\pm 3$.
In semiclassical approximation, the probability of
tunneling between neighboring vacua is determined by instanton configurations.

In the Standard Model, SU(2) instanton lead to an effective 12-fermion interaction
 \[
O_{B+L}=\Pi_{i=1,2,3}q_{Li}q_{Li}q_{Li}l_{Li}\,,
 \]
which describes processes with $\Delta B=\Delta L=3$, such as
\[
u^c+d^c+c^c\to d+2s+2b+t+\nu_e+\nu_\mu+\nu_\tau\,.
 \]
The transition rate is given by
$\Gamma\sim e^{-S_{inst}}\sim {\cal O}(10^{-165})$, where
$S_{inst}$ is instanton action.
Because the rate is extremely small, $B+L$ violating
interactions appear completely negligible in the Standard
Model. However this is not true in a thermal bath, and hence in the primordial Universe.
As emphasized by Kuzmin, Rubakov, and Shaposhnikov, transition between the gauge vacua occurs
not by tunneling but through thermal fluctuations over the barrier.
For temperature $T > E_{sp}$, the suppression in the rate provided by Boltzmann factor
disappears and therefore processes that violate $B+L$ can
occur at a significant rate. In the expanding Universe these processes can be in equilibrium .

\section{The power spectrum of the Gravitational Waves}\label{PWGW}

The function $h({\bf x}, \tau)\equiv h$ appearing in (\ref{h})
satisfies the equation (from Einstein field equation)
 \[
\partial_\tau^2 h +2\frac{\dot a}{a}\partial_\tau h +k^2 h=0\,.
 \]
The mode functions $f_k(\tau)$ obey the minimally coupled
Klein-Gordon equation
 \[
 f_k^{\prime \prime} + \left(k^2 -
 \frac{a^{\prime \prime}}{a}\right)f_k=0\,,
 \]
where ${}^\prime = \partial_\tau$. During de Sitter era, the scale factor $a(\tau)=-1/(H_I \tau)$
where $H_I$ is the Hubble parameter, so that a  solution is
 \[
f_k( \tau)= \frac{e^{-i k \tau}}{\sqrt{2 k}}\left(1- \frac{i}{k
\tau}\right)\,,
\]
which matches the positive frequency "flat space" solutions $e^{-i
k \tau}/\sqrt{2 k}$ in the limit of $k \tau \gg 1$. Substituting
this solution in $ h({\bf x}, \tau)$ (\ref{h}), and using the
canonical commutation relation $[a_k,
a_{k'}^\dagger]=\delta_{kk'}$, we get the standard expression for
two point correlation of gravitational waves generated by
inflation
 \[
 \langle h({\bf x},\tau)h({\bf x},\tau)\rangle^{inf} \equiv \int
\frac{dk} {k}\, (|h_k|^2)^{inf}\,,
 \]
with the spectrum  of gravitational waves given by the scale
invariant form $(|h_k|^2)^{inf}=\frac{4}{\pi} \frac{H_I^2}{m_p^2}$.
Consider now the GWs modes that re-enter the horizon at the
radiation era $a(\tau)\sim \tau$. One finds that the GWs have the
two point correlation fluctuation
 \[
 \langle h({\bf x},\tau)h({\bf x},\tau)\rangle^{rad} \equiv
\int \frac{dk} {k}\, (|h_k|^2)^{rad}\,,
 \]
where
 \[
h^{rad}_k= h^{inf}_k \,\frac{a(T)}{k}\,\frac{T^2\, 1.67\,
 \sqrt{g_*}}{m_P}\,,
 \]
and $g_*=106.7$ is the number of relativistic degrees of freedom which
for the Standard Model

There is a stringent constraint $H_I/M_p < 10^{-5}$ from CMB data
\cite{krauss}. This constraint limits the parameter space of
interactions that can be used for generating the requisite
lepton-asymmetry. In the radiation era, when these modes re-enter
the horizon, the amplitude redshifts by $a^{-1}$ from the time of
re-entry. The reason is that in the radiation era $a(\tau) \sim
\tau$ and the equation for $f_k$ gives plane wave solutions $f_k=
(1/\sqrt{2k})exp(-i k \tau)$. Therefore in the radiation era the
amplitudes of $h$ redshifts as $a^{-1}$. The gravitational waves
inside the horizon in the radiation era will be
$h^{rad}_k= h^{inf}_k \,\frac{a_k}{a(\tau)}=h^{inf}_k
\,\frac{T}{T_k}$,
where $h^{inf}_k$ are the gravitational waves generated by
inflation, $a_k$ and $T_k$ are the scale factor and the
temperature when the modes of wavenumber $k$ entered the horizon
in the radiation era. The horizon entry of mode $k$ occurs when
 \[
 \frac{a_k H_k}{k}= \frac{a(T)\,T\, H_k}{T_k\,k}=1\,,
 \]
where
 \[
H_k= 1.67 \sqrt{g_*} T_k^2 /M_p
 \]
is the Hubble parameter at the time of horizon crossing of the $k$ the mode.
Solving equation for $T_k$ we get
 \[
 T_k= \frac{1}{1.67 \sqrt{g_*}}\frac{k\,M_p}{a(\tau)\,T}\,.
 \]
The amplitude of the gravitational waves of mode $k$ inside the
radiation horizon is (using the equation for $T_k$ and the
previous expression for
 \[
h^{rad} h^{rad}_k= h^{inf}_k \,\frac{a(T)}{k}\,\frac{T^2\, 1.67\, \sqrt{g_*}}{M_p}\,.
 \]
Note that the gravitational wave spectrum inside the radiation era horizon
is no longer scale invariant. The gravitational waves in position
space have the correlation function
\[
\langle h({\bf x},\tau)h({\bf x},\tau)\rangle^{rad} =\int \frac{dk}
{k}\, (h^{rad}_k)^2\,,
\]
and hence for the spin connection $B^0$ generated by the
inflationary gravitational waves in the radiation era, we get
\[
\langle B^0({\bf x},\tau)B^0({\bf x},\tau)\rangle =\int
\frac{dk}{k}\,\left(\frac{k}{a}\right)^2\,(h^{rad}_k)^2
 =\frac{4}{\pi} \left(\frac{H_I}{M_p^2} \, \, T^2\, 1.67\,
\sqrt{g_*}\right)^2 \int_{k_{min}}^{k_{max}} \frac{dk}{k}\,.
\]
The spectrum of spin-connection is scale invariant inside the
radiation horizon. This is significant in that  the lepton
asymmetry generated by this mechanism depends upon the infrared
and ultraviolet scales only logarithmically. The scales outside
the horizon are blue-tilted which means that there will be a scale
dependent anisotropy in the lepton number correlation at two
different space-time points
 \[
\langle \Delta L(r)\Delta L(r^\prime) \rangle \sim A \,k^n\, , n>0\,,
 \]
where
 \[
 \Delta L(r) \equiv L(r)-\bar L\,,
 \]
is the anisotropic deviation from the mean value. Unlike in
the case of CMB, this anisotropy in the lepton number is unlikely
to be accessible to experiments. Nucleosynthesis calculations only
give us an average value at the time of nucleosynthesis (when $T
\sim 1 MeV$). The maximum value of $k$ are for those modes which
leave the de Sitter horizon at the end of inflation. If inflation
is followed by radiation domination era starting with the re-heat
temperature $T_{RH}$ then the maximum value of $k$ in the
radiation era (at temperature $T$) is given by $ k_{max}/a(T)= H_I
\left(\frac{T}{T_{RH}}\right)$. The lower limit of $k$ is
$k_{min}=e^{-N}k_{max}$ which are the modes which left the
de-Sitter horizon in the beginning of inflation ($N$ is the total
e-folding of the scale factor during inflation, $N\simeq 55-70$).
The integration over $k$ then yields just the factor
$\ln(k_{max}/k_{min})=N$.

\end{document}